\documentclass[
    reprint, twocolumn,
    aps,prb,10pt,amsmath,amssymb,
    longbibliography,superscriptaddress,
    showpacs,preprintnumbers]{revtex4-1}

\usepackage{filecontents}

\usepackage[colorlinks=true,citecolor=blue,linkcolor=blue]{hyperref}
\usepackage{oubraces}
\usepackage{bm,bbm,mathbbol}
\usepackage{physics}
\usepackage{mathtools}
\usepackage{amsmath, amsthm, amssymb, amsfonts}
\usepackage{bbold}
\usepackage{esvect}

\usepackage{epsfig}
\usepackage{array}
\usepackage{multirow}
\usepackage{graphicx}
\usepackage[normalem]{ulem}

\usepackage{float}
\usepackage[section]{placeins}
\usepackage{afterpage}

\usepackage{lipsum}

\makeatletter
\newcommand\footnoteref[1]{\protected@xdef\@thefnmark{\ref{#1}}\@footnotemark}
\makeatother

\newcommand{\nnnl}{\nonumber \\}

\def\qe{{\sc Quantum ESPRESSO}}

\newcommand{\mb}[1]{\mathbf{#1}}
\newcommand{\nk}[0]{{n\mathbf{k}}}
\newcommand{\mk}[0]{{m\mathbf{k}}}
\newcommand{\npk}[0]{{n'\mathbf{k}}}

\newcommand{\mkq}[0]{{m\mathbf{k}+\mathbf{q}}}

\newcommand{\qnu}[0]{{\mathbf{q}\nu}}

\newcommand{\veps}[0]{\varepsilon}

\newcommand{\Gka}[0]{{\mb{\Gamma}\kappa\alpha}}
\newcommand{\Gkap}[0]{{\mb{\Gamma}\kappa\alpha'}}
\newcommand{\Gkpap}[0]{{\mb{\Gamma}\kappa'\alpha'}}
\newcommand{\duGka}[0]{{\partial_{\Gka}}}
\newcommand{\duGkap}[0]{{\partial_{\Gkap}}}
\newcommand{\duGkpap}[0]{{\partial_{\Gkpap}}}

\newcommand{\vks}[0]{{\hat{v}_{\rm KS}}}

\newcommand{\riaapprox}[0]{{\stackrel{\text{RIA}}{\approx}}}

\newcommand{\bts}[0]{{\text{BiTlSe$_2$}}}
\newcommand{\dtilde}[0]{{\widetilde{\mathcal{D}}}}

\usepackage{xr}
\makeatletter
\newcommand*{\addFileDependency}[1]{
  \typeout{(#1)}
  \@addtofilelist{#1}
  \IfFileExists{#1}{}{\typeout{No file #1.}}
}
\makeatother

\newcommand*{\myexternaldocument}[1]{
    \externaldocument{#1}
    \addFileDependency{#1.tex}
    \addFileDependency{#1.aux}
}

\myexternaldocument{supp}

\begin{document}

\title{Theory of phonon-induced renormalization of electron wavefunctions}

\author{Jae-Mo Lihm}
\author{Cheol-Hwan Park}
\email{cheolhwan@snu.ac.kr}
\affiliation{Center for Correlated Electron Systems, Institute for Basic Science, Seoul 08826, Korea}
\affiliation{Department of Physics and Astronomy, Seoul National University, Seoul 08826, Korea}
\affiliation{Center for Theoretical Physics, Seoul National University, Seoul 08826, Korea}

\date{\today}

\begin{abstract}
The Allen-Heine-Cardona theory allows us to calculate phonon-induced electron self-energies from first principles without resorting to the adiabatic approximation.
However, this theory has not been able to account for the change of the electron wavefunction, which is crucial if inter-band energy differences are comparable to the phonon-induced electron self-energy as in temperature-driven topological transitions.
Furthermore, for materials without inversion symmetry, even the existence of such topological transitions cannot be investigated using the Allen-Heine-Cardona theory.
Here, we generalize this theory to the renormalization of both the electron energies and wavefunctions.
Our theory can describe both the diagonal and off-diagonal components of the Debye-Waller self-energy in a simple, unified framework.
For demonstration, we calculate the electron-phonon coupling contribution to the temperature-dependent band structure and hidden spin polarization of BiTlSe$_2$ across a topological transition. These quantities can be directly measured.
Our theory opens a new door for studying temperature-induced topological phase transitions in materials both with and without inversion symmetry.
\end{abstract}

\footnotetext[1]{See Supplemental Material, which includes Refs.~\cite{2009GiannozziQE, 2008Eiguren, 2019GonzeABINIT, 2011Gonze, 2013HamannONCVPSP, 2018VanSettenPseudoDojo, 1996PerdewPBE, 2016Antonius, 2015PonceJCP, 2015Verdi, 1997GonzePRB, 2006VanSchilfgaardeGW, 2018Nery, 2019BrownAltvater, 1998ZhangrevPBE, 1976Allen, 2019QueralesFlores, 2017GiustinoRMP}, at [URL will be inserted by publisher] for the detailed derivation of Eq.~\eqref{eq:dtilde_d}, derivation of Eq.~\eqref{eq:dtilde_d} from Eqs.~\eqref{eq:asr_1od} and \eqref{eq:dtilde_od}, numerical tests of Eq.~\eqref{eq:dtilde_od}, the computational details, the convergence study of the electron self-energy, and the temperature dependence of the band gap of \bts.}

\newcommand{\citeSupp}[0]{Note1}

\maketitle

Interactions between electrons and phonons induce a temperature-dependent renormalization of electronic structures~\cite{2017GiustinoRMP}.
The Allen-Heine-Cardona (AHC) theory~\cite{1976Allen,1981Allen,1983Allen} is one of the current state-of-the-art methods to study the effect of electron-phonon coupling (EPC) on electronic structures from first-principles density functional theory (DFT) and density functional perturbation theory (DFPT).
Zero-point renormalization and temperature dependence of the electronic band gap~\cite{2010GiustinoPRL, 2011Gonze, 2014PonceCMS, 2014PoncePRB, 2015Antonius, 2015PonceJCP, 2018Nery, 2019QueralesFlores, 2020Cannuccia}, optical responses~\cite{2008Marini, 2011Cannuccia, 2014Kawai}, and topological properties~\cite{2016Antonius} are being actively investigated with the AHC theory.

In the AHC theory~\cite{1976Allen,1981Allen,1983Allen}, the matrix elements of the second-order derivatives of the electron potential is required to compute the Debye-Waller (DW) contribution to the electron self-energy.
This matrix element is approximated using the rigid-ion approximation (RIA), which assumes that the potential is a sum of atom-centered contributions.
Then, by invoking the translational invariance of the electron eigenvalues, one can calculate the second-order EPC matrix elements from the first-order EPC matrix elements.
However, this method is applicable only to the band-diagonal part of the DW self-energy.
An approximation scheme for the full DW self-energy matrix including off-diagonal components has been absent.

The missing off-diagonal components of the self-energy are essential to describe the hybridization of energy eigenstates~\cite{2017GiustinoRMP,2002LiGW,2013AguileraGW,2016ForsterGW,2018SanchezGW,2019BergesPhonon}.
Thus, the current scope of the first-principles AHC theory is limited in that the EPC-induced change of the electron eigenstate wavefunctions is neglected.
This theoretical limitation even precluded a complete study of electronic structures across an EPC-induced topological transition.
For example, in Ref.~\cite{2016Antonius}, the renormalized electron energy was calculated only at $\Gamma$ where the wavefunction hybridization is forbidden due to the inversion symmetry.
Thus, the evolution of the electronic structure around the topological band-gap inversion could not be investigated.
Furthermore, in materials without inversion symmetry, e.\,g.\,, BiTeI~\cite{2012Bahramy,2013Yang,2014Liu,2017MonserratPRM}, such phonon-induced band-gap inversion occurs not at $\Gamma$ but at generic $k$ points. Hence, the AHC theory cannot even tell the band inversion, let alone the renormalized wavefunctions.
To the best of our knowledge, the role of thermal phonons in renormalizing the electronic structure in the presence of such wavefunction hybridization has never been studied from first principles.

Alternatives of the AHC theory include the frozen phonon method with Monte Carlo integration~\cite{2013Patrick,2016Monserrat_thermal_line,2017MonserratPRM}, the one-shot supercell method~\cite{2015Zacharias,2016Zacharias,2019Zacharias}, and molecular dynamics methods~\cite{2004SalaPIMD,2006RamirezPIMD,2007FranceschettiMD,2008RamirezPIMD}.
However, (i) these methods require calculations of large supercells which are computationally much heavier than DFPT calculations.
Accordingly, the effect of the phonon modes with wave vectors close to the center of the Brillouin zone may not be fully captured.
More importantly, (ii) these methods are intrinsically adiabatic.
Hence, they quantitatively or even qualitatively fail to predict the band gap renormalization of polar semiconductors~\cite{2015PonceJCP,2020Gonze_totalenergy} and cannot describe many-body effects such as the phonon satellites~\cite{2017VerdiPolaron,2018Nery}.

Therefore, it is of great importance to generalize the AHC theory for the study of EPC in a wider variety of materials where the wavefunction hybridization is not negligible. In this paper, we achieve this goal by deriving a simple yet sophisticated expression for the full DW self-energy matrix. This result fills the gap of the AHC theory, rendering it a complete theory of EPC-renormalized electron energies and wavefunctions.
Our formulation also significantly simplifies the computation of the DW self-energy as neither summations over empty states~\cite{1976Allen,1981Allen,1983Allen,2010GiustinoPRL} nor solutions of the Sternheimer equations~\cite{2011Gonze} are required.
The crucial component in this development is the operator generalization of the acoustic sum rule for EPC.
As one of the plentiful applications of our theory, we study the temperature-dependent electronic structures of \bts{} around a topological insulator to normal insulator transition.

In the dynamical AHC theory, the phonon-induced electron self-energy $\Sigma(\omega)$, which is a function of the frequency $\omega$, is written as a sum of the Fan and the DW self-energies~\cite{2017GiustinoRMP,\citeSupp}:
\begin{equation} \label{eq:sigma_def}
    \Sigma_{nn'\mb{k}}(\omega) = \Sigma^{\rm Fan}_{nn'\mb{k}} (\omega) + \Sigma^{\rm DW}_{nn'\mb{k}},
\end{equation}
\begin{align} \label{eq:fan_def}
    &\Sigma^{\rm Fan}_{nn'\mb{k}} (\omega)
    = \frac{1}{N_q}\sum_{\substack{\qnu,m \\ \kappa\alpha\kappa'\alpha'}} \frac{1}{2\omega_{\qnu}}
    [h_{mn}^{\kappa\alpha}(\mb{k},\mb{q})]^* h_{mn'}^{\kappa'\alpha'}(\mb{k},\mb{q})
    \\
    &\times
    U^*_{\kappa\alpha,\nu}(\mb{q}) U_{\kappa'\alpha',\nu}(\mb{q})
    \sum_{\pm} \frac{n_\qnu + [1 \pm (2f_\mkq - 1)]/2}{\omega - \veps_\mkq \pm \omega_\qnu + i\eta},
    \nonumber
\end{align}
\begin{align} \label{eq:dw_def}
    \Sigma^{\rm DW}_{nn'\mb{k}}
    =& \frac{1}{N_q}\sum_{\substack{\qnu \\ \kappa\alpha\kappa'\alpha'}} \frac{1}{2\omega_{\qnu}} \left(n_\qnu + \frac{1}{2} \right) \mathcal{D}_{nn'}^{\kappa\alpha\kappa'\alpha'}(\mb{k},\mb{q}) \\
    &\times U_{\kappa\alpha,\nu}^*(\mb{q}) U_{\kappa'\alpha',\nu}(\mb{q}). \nonumber
\end{align}
Here, $m$, $n$, and $n'$ are the electron band indices, $\mb{k}$ and $\mb{q}$ the electron and phonon crystal momenta, respectively, $U_{\kappa\alpha,\nu}(\mb{q})$ the eigendisplacement of atom
$\kappa$ along Cartesian direction $\alpha$ associated to the phonon mode $\nu$ in units of inverse square root of mass, $\veps_\mkq$ the electron energy, $\omega_\qnu$ the phonon mode energy, $f_\mkq$ and $n_\qnu$ the Fermi-Dirac and Bose-Einstein distribution functions, and $\eta$ a positive infinitesimal that enforces the causality of the EPC. We set $\hbar=1$ throughout this paper.

To compute the Fan and the DW self-energies, one needs to calculate the following two types of matrix elements:
\begin{equation} \label{eq:hcart_def}
    h_{mn}^{\kappa\alpha}(\mb{k},\mb{q})
    = \mel{u_\mkq}{\partial_{\mb{q}\kappa\alpha}\vks}{u_\nk}
\end{equation}
and
\begin{equation} \label{eq:d_def}
    \mathcal{D}_{nn'}^{\kappa\alpha\kappa'\alpha'}(\mb{k},\mb{q})
    = \mel{u_\nk}{\partial_{\mb{-q}\kappa\alpha}\partial_{\mb{q}\kappa'\alpha'}\vks}{u_\npk}.
\end{equation}
Here, $\ket{u_\nk}$ is the periodic part of the electron wavefunction, $\vks$ the Kohn-Sham (KS) potential, and $\partial_{\mb{q}\kappa\alpha}$ the derivative with respect to the monochromatic displacement of atom $\kappa$ along direction $\alpha$ with wave vector $\mb{q}$.

The first-order EPC matrix element $h$ [Eq.~\eqref{eq:hcart_def}] can be evaluated from DFPT.
However, to compute the second-order EPC matrix element $\mathcal{D}$ [Eq.~\eqref{eq:d_def}], we need the second derivatives of the KS potential, which cannot be calculated from the usual first-order DFPT.
Hence, the AHC theory exploits the RIA, assuming that the KS potential is a sum of atom-centered contributions.
Then, mixed derivatives of the KS potential with respect to the displacements of different atoms vanish, allowing one to approximate $\mathcal{D}$ as
\begin{align} \label{eq:d_ria_def}
    \mathcal{D}_{nn'}^{\kappa\alpha\kappa'\alpha'}(\mb{k},\mb{q})
    &\riaapprox \delta_{\kappa,\kappa'} \mel{u_\nk}{\duGka \duGkap \vks}{u_\npk} \nnnl
    &\riaapprox \delta_{\kappa,\kappa'} \sum_{\kappa''} \mel{u_{\nk}}{\duGka \partial_{\mb{\Gamma}\kappa''\alpha'} \vks}{u_{\npk}} \nnnl
    &\equiv \delta_{\kappa,\kappa'} \dtilde_{nn'}^{\kappa\alpha\alpha'}(\mb{k})
\end{align}
with $\delta_{\kappa,\kappa'}$ the Kronecker delta function.
The symbol $\riaapprox$ denotes an approximation that is exact under the RIA.

To compute $\dtilde$ defined in Eq.~\eqref{eq:d_ria_def}, the AHC theory utilizes the translational invariance of electron energies~\cite{1976Allen,2014PoncePRB}.
Let us consider the following operation: a displacement of atom $\kappa$ in every unit cell along direction $\alpha$ by a distance $\xi$ and a subsequent uniform displacement of every atom along direction $\alpha'$ by a distance $\tau$.
We write the KS potential and the energy eigenvalue of the resulting system as $\vks^{\kappa\alpha;\alpha'}(\xi, \tau)$ and $\veps^{\kappa\alpha;\alpha'}_\nk(\xi, \tau)$, respectively.

Now, the translational invariance of the coupled system of electrons and atoms implies that a uniform displacement does not alter the electron eigenenergies:
\begin{equation} \label{eq:veps_tinv}
    \veps^{\kappa\alpha;\alpha'}_\nk(\xi, \tau) = \veps^{\kappa\alpha;\alpha'}_\nk(\xi, 0).
\end{equation}
By expanding both sides of Eq.~\eqref{eq:veps_tinv} with respect to $\xi$ and $\tau$ using perturbation theory, one can relate the diagonal part of $\dtilde$ to the first-order EPC matrix elements~\cite{\citeSupp}:
\begin{equation} \label{eq:dtilde_d}
    \dtilde_{nn}^{\kappa\alpha\alpha'}(\mb{k})
    =- 2 \Re \sum_{\kappa'',m\neq n} 
    \frac{h_{nm}^{\kappa\alpha}(\mb{k},\mb{\Gamma}) h_{mn}^{\kappa''\alpha'}(\mb{k},\mb{\Gamma})}{\veps_\nk - \veps_\mk}.
\end{equation}
This equation is used in the conventional AHC theory to compute the diagonal DW self-energy.

Now, let us consider the full self-energy matrix, including the off-diagonal components. The term `off-diagonal' self-energy is not to be confused with `nondiagonal DW' self-energy~\cite{2011Gonze,2014PoncePRB}, which means the correction to the DW self-energy beyond the RIA.

The Fan self-energy matrix can be computed without any difficulty from Eq.~\eqref{eq:fan_def}.
To compute the DW self-energy matrix, one needs to evaluate the off-diagonal matrix elements of $\dtilde_{nn'}$ [Eq.~\eqref{eq:d_ria_def}].
However, such a quantity cannot be computed within the conventional AHC theory even assuming the RIA, as apparent from Eq.~\eqref{eq:dtilde_d} being limited to the diagonal case.
This limitation originates from the fact that the only nontrivial information we used to derive Eq.~\eqref{eq:dtilde_d} is the translational invariance of the electron energies [Eq.~\eqref{eq:veps_tinv}].

Here, we seek for a statement stronger than Eq.~\eqref{eq:veps_tinv}.
Indeed, the translational invariance of the system gives much more information than the mere invariance of the electron eigenenergies.
When all atoms of the system are uniformly displaced, the electronic Hamiltonian is also uniformly displaced. Thus, we find
\begin{equation} \label{eq:tinv_vks}
    \mel{\mb{r}+\tau\mb{e}_{\alpha'}}{\vks^{\kappa\alpha;\alpha'}(\xi,\tau)}{\mb{r'}+\tau\mb{e}_{\alpha'}}
    = \mel{\mb{r}}{\vks^{\kappa\alpha;\alpha'}(\xi,0)}{\mb{r'}}
\end{equation}
with $\ket{\mb{r}}$ the position basis state of the electron and $\mb{e}_{\alpha'}$ the unit vector along $\alpha'$.

Using the momentum operator $\hat{p}_{\alpha'} = -i{\partial}/{\partial r_{\alpha'}}$,
one can rewrite Eq.~\eqref{eq:tinv_vks} in the operator form:
\begin{equation} \label{eq:vks_p}
    \vks^{\kappa\alpha;\alpha'}(\xi,\tau)
    = \exp(-i \, \tau \,\hat{p}_{\alpha'})\, \vks^{\kappa\alpha;\alpha'}(\xi, 0)\, \exp(i \, \tau\, \hat{p}_{\alpha'}).
\end{equation}
From the coefficients of $\tau$ in the series expansion of Eq.~\eqref{eq:vks_p}, one finds
\begin{equation} \label{eq:asr_1od}
   \sum_{\kappa'} \duGkpap \vks
   = i[\vks, \hat{p}_{\alpha'}].
\end{equation}
From the coefficients of $\xi\tau$, one finds
\begin{equation} \label{eq:asr_2od}
   \sum_{\kappa'} \duGka \duGkpap \vks
   = i[\duGka \vks, \hat{p}_{\alpha'}].
\end{equation}
Here, the derivatives are evaluated at $\xi=\tau=0$, so we omit the superscript $\kappa\alpha;\alpha'$.
Equations~\eqref{eq:asr_1od} and \eqref{eq:asr_2od} are the operator generalization of the acoustic sum rules for EPC.

From Eqs.~\eqref{eq:d_ria_def} and \eqref{eq:asr_2od}, one finds
\begin{equation} \label{eq:dtilde_od}
    \dtilde_{nn'}^{\kappa\alpha\alpha'}(\mb{k})
    = i \mel{u_{\nk}}{[\duGka \vks, \hat{p}_{\alpha'}]}{u_{\npk}}.
\end{equation}
This equation is the main result of this paper.
We have thus generalized the original formula of the AHC theory for the diagonal DW self-energy to the full matrix form.
One can indeed derive the original formula, Eq.~\eqref{eq:dtilde_d}, from our results~\cite{\citeSupp}.
We have numerically tested our formula, Eq.~\eqref{eq:dtilde_od}, against finite-difference calculations and found an excellent agreement~\cite{\citeSupp}.

Using Eq.~\eqref{eq:dtilde_od} as well as Eqs.~\eqref{eq:dw_def} and \eqref{eq:d_ria_def}, one can calculate both the diagonal and off-diagonal components of the DW self-energy matrix on an equal footing.
This result enables the study of the phonon-renormalized electronic structures in presence of considerable wavefunction hybridization, fully within the non-adiabatic perturbative AHC theory.

Another virtue of our theory is that it does not require a summation over unoccupied bands, in contrast to the original expression Eq.~\eqref{eq:dtilde_d}~\cite{1976Allen,1981Allen,1983Allen,2010GiustinoPRL}, or solutions of the Sternheimer equations~\cite{2011Gonze}. The computation of the DW self-energy is thus considerably simplified.

\begin{figure}
\includegraphics[width=0.9\columnwidth]{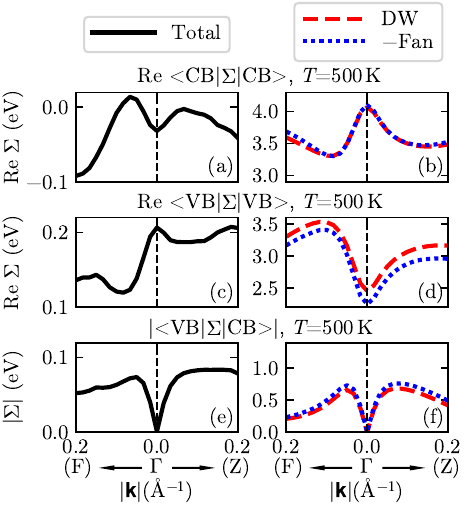}
\caption{
The total self-energy and the DW and Fan contributions at $T=500\,K$ of the (a-b) conduction band, (c-d) valence band, and (e-f) off-diagonal elements.
For the Fan self-energy, we show $- \Re \Sigma^{\rm Fan}$ in (b) and (d), and $\abs{\Sigma^{\rm Fan}}$ in (f).
The vertical axes of the three left and three right panels have the same scales among themselves.
}
\label{fig:fandwtot}
\end{figure}

\begin{figure*}
\includegraphics[width=1.0\textwidth]{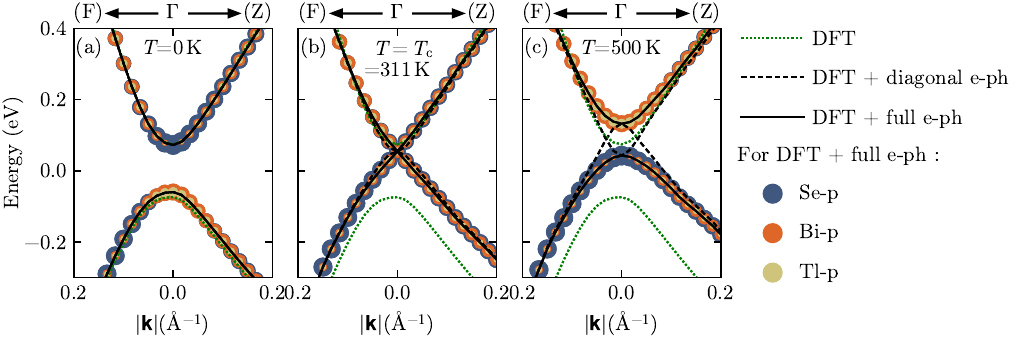}
\caption{
Temperature dependence of the band structure of \bts{}. The size of the colored circles are proportional to the projection of the electron wavefunctions onto the atomic orbitals.
}
\label{fig:band}
\end{figure*}

\begin{figure*}
\includegraphics[width=1.0\textwidth]{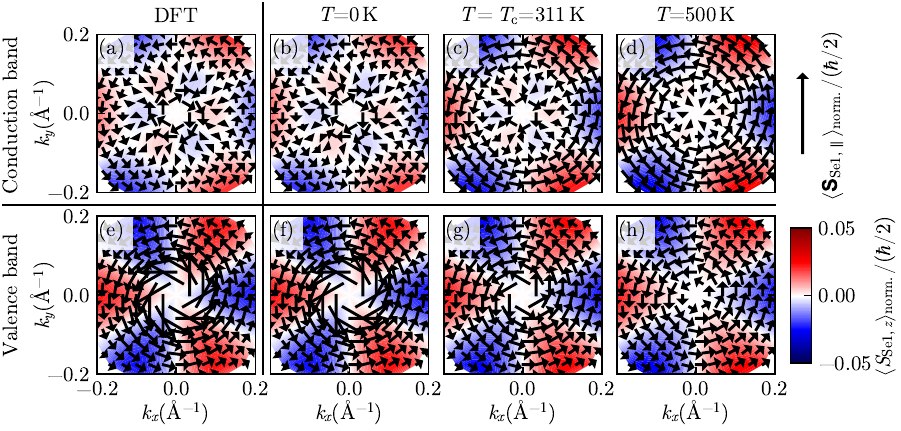}
\caption{
In-plane and out-of-plane components of the hidden spin polarization of the Se1 atom on the $k_z=0$ plane.
}
\label{fig:spol}
\end{figure*}

\footnotetext[5]{The custom code developed in this work may be made available in a later release of the \qe\ package after discussions with the \qe\ core developers.}
\newcommand{\citeNoteCode}[0]{Note5}

We apply our theory to the temperature-dependent electronic structure of \bts{}. 
\bts{} exhibits an EPC-induced topological transition~\cite{2016Antonius}.
DFT calculations of \bts{} find a topological insulator state with an inverted gap at $\Gamma$~\cite{2011EremeevPRB}, as have been experimentally observed~\cite{2010SatoPRL, 2010KurodaPRL, 2010ChenPRL}.
When the EPC is included, \bts{} is driven to a normal insulator at temperatures higher than the critical temperature~\cite{2016Antonius}.
Using our generalization of the AHC theory, we study the evolution of the band structures and wavefunctions around this topological transition~\cite{\citeSupp, \citeNoteCode}.
In calculating the temperature dependence, we neglected the effect of thermal expansion as done in Ref.~\cite{2016Antonius}.

Figure~\ref{fig:fandwtot} shows the decomposition of the total self-energy into the Fan and DW contributions.
We find that the two contributions have similar magnitudes and opposite signs.
Even for the off-diagonal components which are in general complex, the Fan and DW self-energies have approximately opposite phase angles.
The magnitude of the total self-energy is thus much smaller than the magnitudes of the Fan and DW contributions.
This trend is in accordance with previous analytical~\cite{1976Allen} and numerical~\cite{2019QueralesFlores} studies which also reported a cancellation between the Fan and DW self-energies in direct narrow-gap semiconductors.
This {\it almost} cancellation between the Fan and DW self-energies for both diagonal and off-diagonal matrix elements clearly shows the importance of our new method.

Figure~\ref{fig:band} shows the temperature dependence of the band structure and the orbital character of \bts.
Figure~\ref{fig:band}(a) corresponds to a low-temperature topological insulator phase and shows the inversion of orbital character around $\Gamma$.
Figure~\ref{fig:band}(b) shows a gapless linear dispersion of the renormalized electron bands at $T=T_{\rm c}$~\cite{\citeSupp}.
Figure~\ref{fig:band}(c) shows that at temperatures above $T_{\rm c}$, the orbital characters of the valence and the conduction bands change smoothly, indicating a trivial insulator state.
The topological transition can also be identified from the parity of the valence band wavefunction at the $\Gamma$ point~\cite{2007Fu}.

The changes in the band structure and the orbital characters can be correctly calculated only when the full self-energy matrix is taken into account.
The electron energies computed using only the diagonal part of the self-energy show unphysical features such as the quadratic band touching [Fig.~\ref{fig:band}(b)] and band crossings [Fig.~\ref{fig:band}(c)].

As another example of the physical quantity affected by the phonon-induced wavefunction hybridization, we calculate the hidden spin polarization.
The structure of \bts{} has inversion symmetry, so the electron eigenstates do not have any spin polarization.
Still, local spin polarization can exist at atoms that are not located at an inversion center.
Such local spin polarization is termed ``hidden spin polarization''~\cite{2014ZhangHSPth}
and has been experimentally measured~\cite{2014RileyHSPexp, 2017YaoHSPOL, 2017RazzoliHSPOL, 2018ChengHSPOL, 2020OlivaHSPOL, 2020TuHSPOL}.

Figure~\ref{fig:spol} shows the normalized hidden spin polarization of one of the two selenium atoms in the unit cell (Se1)~\cite{\citeSupp}.
The normalized hidden spin polarization is defined as
\begin{equation}
    \expval{\mb{S}_{\rm Se1}}_{\rm norm.}
    = \mel{\psi_\nk}{\hat{\mb{S}} \hat{P}_{\rm Se1}}{\psi_\nk} / \mel{\psi_\nk}{\hat{P}_{\rm Se1}}{\psi_\nk}
\end{equation}
where $\hat{\mb{S}}$ is the spin angular momentum operator and $\hat{P}_{\rm Se1}$ is the projection operator to the valence $p$ orbitals of Se1 atom.

From the DFT calculation [Fig.~\ref{fig:spol}(a)] and also after the inclusion of zero-point renormalization [Fig.~\ref{fig:spol}(b)], we find that the hidden spin texture of the conduction band is flipped as the $k$ point moves away from $\Gamma$.
This finding indicates an inverted gap at $\Gamma$ at low temperatures.
As the temperature increases, the hidden spin polarization near $\Gamma$ is reversed.
At $T=500\,K$, smoothly varying hidden spin textures are found for both the conduction and the valence bands [Figs.~\ref{fig:spol}(d) and~\ref{fig:spol}(h)].
We emphasize that the inclusion of the off-diagonal components of the self-energy matrix is essential to study any kind of wavefunction-dependent quantities, of which hidden spin polarization is only a single example.

The range of systems with significant phonon-induced wavefunction hybridization is far more diverse than that studied in this work.
For example, the appearance and disappearance of topologically-protected surface states after a temperature-induced topological transition of the bulk is a result of significant wavefunction hybridization.
Also, the consideration of the full self-energy matrix is necessary to study the EPC-induced topological transitions of non-centrosymmetric systems such as BiTeI~\cite{2012Bahramy,2013Yang,2014Liu,2017MonserratPRM}, as the gap closes at generic $k$ points where the symmetry allows interband hybridization.
Our theory also provides the temperature-depenent electronic wavefunctions which are needed for the calculation of the topological $\mathbb{Z}_2$ invariant of non-centrosymmetric systems~\cite{Soluyanov2011Z2,Yu2011Z2,Ringel2011Z2,Fukui2007Z2}.

In conclusion, we derive a simple expression for the full DW self-energy matrix using the operator generalization of the acoustic sum rule for EPC.
This result generalizes the AHC theory to deal with the off-diagonal components of the electron self-energy and the EPC-induced wavefunction hybridization.
We demonstrate our formalism by calculating the EPC contribution to the temperature dependence of the band structures, orbital characters, and hidden spin polarization of \bts{}.
We find that the consideration of the full self-energy matrix is essential for a complete understanding of the gap inversion at the topological transition.
Our work paves the way for the study of temperature-dependent electronic structures of a far wider variety of systems, fully within the dynamical AHC theory.

\begin{acknowledgments}
This work was supported by the Institute for Basic Science (No. IBSR009-D1) and by the Creative-Pioneering Research Program through Seoul National University.
\end{acknowledgments}

\makeatletter\@input{xy.tex}\makeatother

\bibliography{main}

%apsrev4-2.bst 2019-01-14 (MD) hand-edited version of apsrev4-1.bst
%Control: key (0)
%Control: author (8) initials jnrlst
%Control: editor formatted (1) identically to author
%Control: production of article title (0) allowed
%Control: page (0) single
%Control: year (1) truncated
%Control: production of eprint (0) enabled
\begin{thebibliography}{66}%
\makeatletter
\providecommand \@ifxundefined [1]{%
 \@ifx{#1\undefined}
}%
\providecommand \@ifnum [1]{%
 \ifnum #1\expandafter \@firstoftwo
 \else \expandafter \@secondoftwo
 \fi
}%
\providecommand \@ifx [1]{%
 \ifx #1\expandafter \@firstoftwo
 \else \expandafter \@secondoftwo
 \fi
}%
\providecommand \natexlab [1]{#1}%
\providecommand \enquote  [1]{``#1''}%
\providecommand \bibnamefont  [1]{#1}%
\providecommand \bibfnamefont [1]{#1}%
\providecommand \citenamefont [1]{#1}%
\providecommand \href@noop [0]{\@secondoftwo}%
\providecommand \href [0]{\begingroup \@sanitize@url \@href}%
\providecommand \@href[1]{\@@startlink{#1}\@@href}%
\providecommand \@@href[1]{\endgroup#1\@@endlink}%
\providecommand \@sanitize@url [0]{\catcode `\\12\catcode `\$12\catcode
  `\&12\catcode `\#12\catcode `\^12\catcode `\_12\catcode `\%12\relax}%
\providecommand \@@startlink[1]{}%
\providecommand \@@endlink[0]{}%
\providecommand \url  [0]{\begingroup\@sanitize@url \@url }%
\providecommand \@url [1]{\endgroup\@href {#1}{\urlprefix }}%
\providecommand \urlprefix  [0]{URL }%
\providecommand \Eprint [0]{\href }%
\providecommand \doibase [0]{https://doi.org/}%
\providecommand \selectlanguage [0]{\@gobble}%
\providecommand \bibinfo  [0]{\@secondoftwo}%
\providecommand \bibfield  [0]{\@secondoftwo}%
\providecommand \translation [1]{[#1]}%
\providecommand \BibitemOpen [0]{}%
\providecommand \bibitemStop [0]{}%
\providecommand \bibitemNoStop [0]{.\EOS\space}%
\providecommand \EOS [0]{\spacefactor3000\relax}%
\providecommand \BibitemShut  [1]{\csname bibitem#1\endcsname}%
\let\auto@bib@innerbib\@empty
%</preamble>
\bibitem [{\citenamefont {Giustino}(2017)}]{2017GiustinoRMP}%
  \BibitemOpen
  \bibfield  {author} {\bibinfo {author} {\bibfnamefont {F.}~\bibnamefont
  {Giustino}},\ }\bibfield  {title} {\bibinfo {title} {Electron-phonon
  interactions from first principles},\ }\href
  {https://doi.org/10.1103/RevModPhys.89.015003} {\bibfield  {journal}
  {\bibinfo  {journal} {Reviews of Modern Physics}\ }\textbf {\bibinfo {volume}
  {89}},\ \bibinfo {pages} {015003} (\bibinfo {year} {2017})}\BibitemShut
  {NoStop}%
\bibitem [{\citenamefont {Allen}\ and\ \citenamefont
  {Heine}(1976)}]{1976Allen}%
  \BibitemOpen
  \bibfield  {author} {\bibinfo {author} {\bibfnamefont {P.~B.}\ \bibnamefont
  {Allen}}\ and\ \bibinfo {author} {\bibfnamefont {V.}~\bibnamefont {Heine}},\
  }\bibfield  {title} {\bibinfo {title} {Theory of the temperature dependence
  of electronic band structures},\ }\href
  {https://doi.org/10.1088/0022-3719/9/12/013} {\bibfield  {journal} {\bibinfo
  {journal} {Journal of Physics C: Solid State Physics}\ }\textbf {\bibinfo
  {volume} {9}},\ \bibinfo {pages} {2305} (\bibinfo {year} {1976})}\BibitemShut
  {NoStop}%
\bibitem [{\citenamefont {Allen}\ and\ \citenamefont
  {Cardona}(1981)}]{1981Allen}%
  \BibitemOpen
  \bibfield  {author} {\bibinfo {author} {\bibfnamefont {P.~B.}\ \bibnamefont
  {Allen}}\ and\ \bibinfo {author} {\bibfnamefont {M.}~\bibnamefont
  {Cardona}},\ }\bibfield  {title} {\bibinfo {title} {Theory of the temperature
  dependence of the direct gap of germanium},\ }\href
  {https://doi.org/10.1103/PhysRevB.23.1495} {\bibfield  {journal} {\bibinfo
  {journal} {Physical Review B}\ }\textbf {\bibinfo {volume} {23}},\ \bibinfo
  {pages} {1495} (\bibinfo {year} {1981})}\BibitemShut {NoStop}%
\bibitem [{\citenamefont {Allen}\ and\ \citenamefont
  {Cardona}(1983)}]{1983Allen}%
  \BibitemOpen
  \bibfield  {author} {\bibinfo {author} {\bibfnamefont {P.~B.}\ \bibnamefont
  {Allen}}\ and\ \bibinfo {author} {\bibfnamefont {M.}~\bibnamefont
  {Cardona}},\ }\bibfield  {title} {\bibinfo {title} {Temperature dependence of
  the direct gap of {Si} and {Ge}},\ }\href
  {https://doi.org/10.1103/PhysRevB.27.4760} {\bibfield  {journal} {\bibinfo
  {journal} {Phys. Rev. B}\ }\textbf {\bibinfo {volume} {27}},\ \bibinfo
  {pages} {4760} (\bibinfo {year} {1983})}\BibitemShut {NoStop}%
\bibitem [{\citenamefont {Giustino}\ \emph {et~al.}(2010)\citenamefont
  {Giustino}, \citenamefont {Louie},\ and\ \citenamefont
  {Cohen}}]{2010GiustinoPRL}%
  \BibitemOpen
  \bibfield  {author} {\bibinfo {author} {\bibfnamefont {F.}~\bibnamefont
  {Giustino}}, \bibinfo {author} {\bibfnamefont {S.~G.}\ \bibnamefont
  {Louie}},\ and\ \bibinfo {author} {\bibfnamefont {M.~L.}\ \bibnamefont
  {Cohen}},\ }\bibfield  {title} {\bibinfo {title} {Electron-phonon
  renormalization of the direct band gap of diamond},\ }\href
  {https://doi.org/10.1103/PhysRevLett.105.265501} {\bibfield  {journal}
  {\bibinfo  {journal} {Physical Review Letters}\ }\textbf {\bibinfo {volume}
  {105}},\ \bibinfo {pages} {265501} (\bibinfo {year} {2010})}\BibitemShut
  {NoStop}%
\bibitem [{\citenamefont {Gonze}\ \emph {et~al.}(2011)\citenamefont {Gonze},
  \citenamefont {Boulanger},\ and\ \citenamefont {Côté}}]{2011Gonze}%
  \BibitemOpen
  \bibfield  {author} {\bibinfo {author} {\bibfnamefont {X.}~\bibnamefont
  {Gonze}}, \bibinfo {author} {\bibfnamefont {P.}~\bibnamefont {Boulanger}},\
  and\ \bibinfo {author} {\bibfnamefont {M.}~\bibnamefont {Côté}},\
  }\bibfield  {title} {\bibinfo {title} {Theoretical approaches to the
  temperature and zero-point motion effects on the electronic band structure},\
  }\href {https://doi.org/10.1002/andp.201000100} {\bibfield  {journal}
  {\bibinfo  {journal} {Annalen der Physik}\ }\textbf {\bibinfo {volume}
  {523}},\ \bibinfo {pages} {168} (\bibinfo {year} {2011})}\BibitemShut
  {NoStop}%
\bibitem [{\citenamefont {Poncé}\ \emph
  {et~al.}(2014{\natexlab{a}})\citenamefont {Poncé}, \citenamefont {Antonius},
  \citenamefont {Boulanger}, \citenamefont {Cannuccia}, \citenamefont {Marini},
  \citenamefont {Côté},\ and\ \citenamefont {Gonze}}]{2014PonceCMS}%
  \BibitemOpen
  \bibfield  {author} {\bibinfo {author} {\bibfnamefont {S.}~\bibnamefont
  {Poncé}}, \bibinfo {author} {\bibfnamefont {G.}~\bibnamefont {Antonius}},
  \bibinfo {author} {\bibfnamefont {P.}~\bibnamefont {Boulanger}}, \bibinfo
  {author} {\bibfnamefont {E.}~\bibnamefont {Cannuccia}}, \bibinfo {author}
  {\bibfnamefont {A.}~\bibnamefont {Marini}}, \bibinfo {author} {\bibfnamefont
  {M.}~\bibnamefont {Côté}},\ and\ \bibinfo {author} {\bibfnamefont
  {X.}~\bibnamefont {Gonze}},\ }\bibfield  {title} {\bibinfo {title}
  {Verification of first-principles codes: {Comparison} of total energies,
  phonon frequencies, electron–phonon coupling and zero-point motion
  correction to the gap between {ABINIT} and {QE}/{Yambo}},\ }\href
  {https://doi.org/10.1016/j.commatsci.2013.11.031} {\bibfield  {journal}
  {\bibinfo  {journal} {Computational Materials Science}\ }\textbf {\bibinfo
  {volume} {83}},\ \bibinfo {pages} {341} (\bibinfo {year}
  {2014}{\natexlab{a}})}\BibitemShut {NoStop}%
\bibitem [{\citenamefont {Poncé}\ \emph
  {et~al.}(2014{\natexlab{b}})\citenamefont {Poncé}, \citenamefont {Antonius},
  \citenamefont {Gillet}, \citenamefont {Boulanger}, \citenamefont
  {Laflamme~Janssen}, \citenamefont {Marini}, \citenamefont {Côté},\ and\
  \citenamefont {Gonze}}]{2014PoncePRB}%
  \BibitemOpen
  \bibfield  {author} {\bibinfo {author} {\bibfnamefont {S.}~\bibnamefont
  {Poncé}}, \bibinfo {author} {\bibfnamefont {G.}~\bibnamefont {Antonius}},
  \bibinfo {author} {\bibfnamefont {Y.}~\bibnamefont {Gillet}}, \bibinfo
  {author} {\bibfnamefont {P.}~\bibnamefont {Boulanger}}, \bibinfo {author}
  {\bibfnamefont {J.}~\bibnamefont {Laflamme~Janssen}}, \bibinfo {author}
  {\bibfnamefont {A.}~\bibnamefont {Marini}}, \bibinfo {author} {\bibfnamefont
  {M.}~\bibnamefont {Côté}},\ and\ \bibinfo {author} {\bibfnamefont
  {X.}~\bibnamefont {Gonze}},\ }\bibfield  {title} {\bibinfo {title}
  {Temperature dependence of electronic eigenenergies in the adiabatic harmonic
  approximation},\ }\href {https://doi.org/10.1103/PhysRevB.90.214304}
  {\bibfield  {journal} {\bibinfo  {journal} {Physical Review B}\ }\textbf
  {\bibinfo {volume} {90}},\ \bibinfo {pages} {214304} (\bibinfo {year}
  {2014}{\natexlab{b}})}\BibitemShut {NoStop}%
\bibitem [{\citenamefont {Antonius}\ \emph {et~al.}(2015)\citenamefont
  {Antonius}, \citenamefont {Poncé}, \citenamefont {Lantagne-Hurtubise},
  \citenamefont {Auclair}, \citenamefont {Gonze},\ and\ \citenamefont
  {Côté}}]{2015Antonius}%
  \BibitemOpen
  \bibfield  {author} {\bibinfo {author} {\bibfnamefont {G.}~\bibnamefont
  {Antonius}}, \bibinfo {author} {\bibfnamefont {S.}~\bibnamefont {Poncé}},
  \bibinfo {author} {\bibfnamefont {E.}~\bibnamefont {Lantagne-Hurtubise}},
  \bibinfo {author} {\bibfnamefont {G.}~\bibnamefont {Auclair}}, \bibinfo
  {author} {\bibfnamefont {X.}~\bibnamefont {Gonze}},\ and\ \bibinfo {author}
  {\bibfnamefont {M.}~\bibnamefont {Côté}},\ }\bibfield  {title} {\bibinfo
  {title} {Dynamical and anharmonic effects on the electron-phonon coupling and
  the zero-point renormalization of the electronic structure},\ }\href
  {https://doi.org/10.1103/PhysRevB.92.085137} {\bibfield  {journal} {\bibinfo
  {journal} {Physical Review B}\ }\textbf {\bibinfo {volume} {92}},\ \bibinfo
  {pages} {085137} (\bibinfo {year} {2015})}\BibitemShut {NoStop}%
\bibitem [{\citenamefont {Poncé}\ \emph {et~al.}(2015)\citenamefont {Poncé},
  \citenamefont {Gillet}, \citenamefont {Laflamme~Janssen}, \citenamefont
  {Marini}, \citenamefont {Verstraete},\ and\ \citenamefont
  {Gonze}}]{2015PonceJCP}%
  \BibitemOpen
  \bibfield  {author} {\bibinfo {author} {\bibfnamefont {S.}~\bibnamefont
  {Poncé}}, \bibinfo {author} {\bibfnamefont {Y.}~\bibnamefont {Gillet}},
  \bibinfo {author} {\bibfnamefont {J.}~\bibnamefont {Laflamme~Janssen}},
  \bibinfo {author} {\bibfnamefont {A.}~\bibnamefont {Marini}}, \bibinfo
  {author} {\bibfnamefont {M.}~\bibnamefont {Verstraete}},\ and\ \bibinfo
  {author} {\bibfnamefont {X.}~\bibnamefont {Gonze}},\ }\bibfield  {title}
  {\bibinfo {title} {Temperature dependence of the electronic structure of
  semiconductors and insulators},\ }\href {https://doi.org/10.1063/1.4927081}
  {\bibfield  {journal} {\bibinfo  {journal} {The Journal of Chemical Physics}\
  }\textbf {\bibinfo {volume} {143}},\ \bibinfo {pages} {102813} (\bibinfo
  {year} {2015})}\BibitemShut {NoStop}%
\bibitem [{\citenamefont {Nery}\ \emph {et~al.}(2018)\citenamefont {Nery},
  \citenamefont {Allen}, \citenamefont {Antonius}, \citenamefont {Reining},
  \citenamefont {Miglio},\ and\ \citenamefont {Gonze}}]{2018Nery}%
  \BibitemOpen
  \bibfield  {author} {\bibinfo {author} {\bibfnamefont {J.~P.}\ \bibnamefont
  {Nery}}, \bibinfo {author} {\bibfnamefont {P.~B.}\ \bibnamefont {Allen}},
  \bibinfo {author} {\bibfnamefont {G.}~\bibnamefont {Antonius}}, \bibinfo
  {author} {\bibfnamefont {L.}~\bibnamefont {Reining}}, \bibinfo {author}
  {\bibfnamefont {A.}~\bibnamefont {Miglio}},\ and\ \bibinfo {author}
  {\bibfnamefont {X.}~\bibnamefont {Gonze}},\ }\bibfield  {title} {\bibinfo
  {title} {Quasiparticles and phonon satellites in spectral functions of
  semiconductors and insulators: {Cumulants} applied to the full
  first-principles theory and the {Fr\"ohlich} polaron},\ }\href
  {https://doi.org/10.1103/PhysRevB.97.115145} {\bibfield  {journal} {\bibinfo
  {journal} {Physical Review B}\ }\textbf {\bibinfo {volume} {97}},\ \bibinfo
  {pages} {115145} (\bibinfo {year} {2018})}\BibitemShut {NoStop}%
\bibitem [{\citenamefont {Querales-Flores}\ \emph {et~al.}(2019)\citenamefont
  {Querales-Flores}, \citenamefont {Cao}, \citenamefont {Fahy},\ and\
  \citenamefont {Savić}}]{2019QueralesFlores}%
  \BibitemOpen
  \bibfield  {author} {\bibinfo {author} {\bibfnamefont {J.~D.}\ \bibnamefont
  {Querales-Flores}}, \bibinfo {author} {\bibfnamefont {J.}~\bibnamefont
  {Cao}}, \bibinfo {author} {\bibfnamefont {S.}~\bibnamefont {Fahy}},\ and\
  \bibinfo {author} {\bibfnamefont {I.}~\bibnamefont {Savić}},\ }\bibfield
  {title} {\bibinfo {title} {Temperature effects on the electronic band
  structure of {PbTe} from first principles},\ }\href
  {https://doi.org/10.1103/PhysRevMaterials.3.055405} {\bibfield  {journal}
  {\bibinfo  {journal} {Physical Review Materials}\ }\textbf {\bibinfo {volume}
  {3}},\ \bibinfo {pages} {055405} (\bibinfo {year} {2019})}\BibitemShut
  {NoStop}%
\bibitem [{\citenamefont {Cannuccia}\ and\ \citenamefont
  {Gali}(2020)}]{2020Cannuccia}%
  \BibitemOpen
  \bibfield  {author} {\bibinfo {author} {\bibfnamefont {E.}~\bibnamefont
  {Cannuccia}}\ and\ \bibinfo {author} {\bibfnamefont {A.}~\bibnamefont
  {Gali}},\ }\bibfield  {title} {\bibinfo {title} {Thermal evolution of silicon
  carbide electronic bands},\ }\href
  {https://doi.org/10.1103/PhysRevMaterials.4.014601} {\bibfield  {journal}
  {\bibinfo  {journal} {Physical Review Materials}\ }\textbf {\bibinfo {volume}
  {4}},\ \bibinfo {pages} {014601} (\bibinfo {year} {2020})}\BibitemShut
  {NoStop}%
\bibitem [{\citenamefont {Marini}(2008)}]{2008Marini}%
  \BibitemOpen
  \bibfield  {author} {\bibinfo {author} {\bibfnamefont {A.}~\bibnamefont
  {Marini}},\ }\bibfield  {title} {\bibinfo {title} {\textit{Ab Initio}
  finite-temperature excitons},\ }\href
  {https://doi.org/10.1103/PhysRevLett.101.106405} {\bibfield  {journal}
  {\bibinfo  {journal} {Physical Review Letters}\ }\textbf {\bibinfo {volume}
  {101}},\ \bibinfo {pages} {106405} (\bibinfo {year} {2008})}\BibitemShut
  {NoStop}%
\bibitem [{\citenamefont {Cannuccia}\ and\ \citenamefont
  {Marini}(2011)}]{2011Cannuccia}%
  \BibitemOpen
  \bibfield  {author} {\bibinfo {author} {\bibfnamefont {E.}~\bibnamefont
  {Cannuccia}}\ and\ \bibinfo {author} {\bibfnamefont {A.}~\bibnamefont
  {Marini}},\ }\bibfield  {title} {\bibinfo {title} {Effect of the quantum
  zero-point atomic motion on the optical and electronic properties of diamond
  and trans-polyacetylene},\ }\href
  {https://doi.org/10.1103/PhysRevLett.107.255501} {\bibfield  {journal}
  {\bibinfo  {journal} {Physical Review Letters}\ }\textbf {\bibinfo {volume}
  {107}},\ \bibinfo {pages} {255501} (\bibinfo {year} {2011})}\BibitemShut
  {NoStop}%
\bibitem [{\citenamefont {Kawai}\ \emph {et~al.}(2014)\citenamefont {Kawai},
  \citenamefont {Yamashita}, \citenamefont {Cannuccia},\ and\ \citenamefont
  {Marini}}]{2014Kawai}%
  \BibitemOpen
  \bibfield  {author} {\bibinfo {author} {\bibfnamefont {H.}~\bibnamefont
  {Kawai}}, \bibinfo {author} {\bibfnamefont {K.}~\bibnamefont {Yamashita}},
  \bibinfo {author} {\bibfnamefont {E.}~\bibnamefont {Cannuccia}},\ and\
  \bibinfo {author} {\bibfnamefont {A.}~\bibnamefont {Marini}},\ }\bibfield
  {title} {\bibinfo {title} {Electron-electron and electron-phonon correlation
  effects on the finite-temperature electronic and optical properties of
  zinc-blende {GaN}},\ }\href {https://doi.org/10.1103/PhysRevB.89.085202}
  {\bibfield  {journal} {\bibinfo  {journal} {Physical Review B}\ }\textbf
  {\bibinfo {volume} {89}},\ \bibinfo {pages} {085202} (\bibinfo {year}
  {2014})}\BibitemShut {NoStop}%
\bibitem [{\citenamefont {Antonius}\ and\ \citenamefont
  {Louie}(2016)}]{2016Antonius}%
  \BibitemOpen
  \bibfield  {author} {\bibinfo {author} {\bibfnamefont {G.}~\bibnamefont
  {Antonius}}\ and\ \bibinfo {author} {\bibfnamefont {S.~G.}\ \bibnamefont
  {Louie}},\ }\bibfield  {title} {\bibinfo {title} {Temperature-induced
  topological phase transitions: Promoted versus suppressed nontrivial
  topology},\ }\href {https://doi.org/10.1103/PhysRevLett.117.246401}
  {\bibfield  {journal} {\bibinfo  {journal} {Physical Review Letters}\
  }\textbf {\bibinfo {volume} {117}},\ \bibinfo {pages} {246401} (\bibinfo
  {year} {2016})}\BibitemShut {NoStop}%
\bibitem [{\citenamefont {Li}\ \emph {et~al.}(2002)\citenamefont {Li},
  \citenamefont {Rignanese}, \citenamefont {Chang}, \citenamefont {Blase},\
  and\ \citenamefont {Louie}}]{2002LiGW}%
  \BibitemOpen
  \bibfield  {author} {\bibinfo {author} {\bibfnamefont {J.-L.}\ \bibnamefont
  {Li}}, \bibinfo {author} {\bibfnamefont {G.-M.}\ \bibnamefont {Rignanese}},
  \bibinfo {author} {\bibfnamefont {E.~K.}\ \bibnamefont {Chang}}, \bibinfo
  {author} {\bibfnamefont {X.}~\bibnamefont {Blase}},\ and\ \bibinfo {author}
  {\bibfnamefont {S.~G.}\ \bibnamefont {Louie}},\ }\bibfield  {title} {\bibinfo
  {title} {{$GW$} study of the metal-insulator transition of bcc hydrogen},\
  }\href {https://doi.org/10.1103/PhysRevB.66.035102} {\bibfield  {journal}
  {\bibinfo  {journal} {Physical Review B}\ }\textbf {\bibinfo {volume} {66}},\
  \bibinfo {pages} {035102} (\bibinfo {year} {2002})}\BibitemShut {NoStop}%
\bibitem [{\citenamefont {Aguilera}\ \emph {et~al.}(2013)\citenamefont
  {Aguilera}, \citenamefont {Friedrich}, \citenamefont {Bihlmayer},\ and\
  \citenamefont {Blügel}}]{2013AguileraGW}%
  \BibitemOpen
  \bibfield  {author} {\bibinfo {author} {\bibfnamefont {I.}~\bibnamefont
  {Aguilera}}, \bibinfo {author} {\bibfnamefont {C.}~\bibnamefont {Friedrich}},
  \bibinfo {author} {\bibfnamefont {G.}~\bibnamefont {Bihlmayer}},\ and\
  \bibinfo {author} {\bibfnamefont {S.}~\bibnamefont {Blügel}},\ }\bibfield
  {title} {\bibinfo {title} {{$GW$} study of topological insulators
  {Bi$_2$Se$_3$}, {Bi$_2$Te$_3$}, and {Sb$_2$Te$_3$}: Beyond the perturbative
  one-shot approach},\ }\href {https://doi.org/10.1103/PhysRevB.88.045206}
  {\bibfield  {journal} {\bibinfo  {journal} {Physical Review B}\ }\textbf
  {\bibinfo {volume} {88}},\ \bibinfo {pages} {045206} (\bibinfo {year}
  {2013})}\BibitemShut {NoStop}%
\bibitem [{\citenamefont {Förster}\ \emph {et~al.}(2016)\citenamefont
  {Förster}, \citenamefont {Krüger},\ and\ \citenamefont
  {Rohlfing}}]{2016ForsterGW}%
  \BibitemOpen
  \bibfield  {author} {\bibinfo {author} {\bibfnamefont {T.}~\bibnamefont
  {Förster}}, \bibinfo {author} {\bibfnamefont {P.}~\bibnamefont {Krüger}},\
  and\ \bibinfo {author} {\bibfnamefont {M.}~\bibnamefont {Rohlfing}},\
  }\bibfield  {title} {\bibinfo {title} {{$GW$} calculations for {Bi$_2$Te$_3$}
  and {Sb$_2$Te$_3$} thin films: Electronic and topological properties},\
  }\href {https://doi.org/10.1103/PhysRevB.93.205442} {\bibfield  {journal}
  {\bibinfo  {journal} {Physical Review B}\ }\textbf {\bibinfo {volume} {93}},\
  \bibinfo {pages} {205442} (\bibinfo {year} {2016})}\BibitemShut {NoStop}%
\bibitem [{\citenamefont {Sánchez-Barriga}\ \emph {et~al.}(2018)\citenamefont
  {Sánchez-Barriga}, \citenamefont {Aguilera}, \citenamefont {Yashina},
  \citenamefont {Tsukanova}, \citenamefont {Freyse}, \citenamefont {Chaika},
  \citenamefont {Callaert}, \citenamefont {Abakumov}, \citenamefont
  {Hadermann}, \citenamefont {Varykhalov}, \citenamefont {Rienks},
  \citenamefont {Bihlmayer}, \citenamefont {Blügel},\ and\ \citenamefont
  {Rader}}]{2018SanchezGW}%
  \BibitemOpen
  \bibfield  {author} {\bibinfo {author} {\bibfnamefont {J.}~\bibnamefont
  {Sánchez-Barriga}}, \bibinfo {author} {\bibfnamefont {I.}~\bibnamefont
  {Aguilera}}, \bibinfo {author} {\bibfnamefont {L.~V.}\ \bibnamefont
  {Yashina}}, \bibinfo {author} {\bibfnamefont {D.~Y.}\ \bibnamefont
  {Tsukanova}}, \bibinfo {author} {\bibfnamefont {F.}~\bibnamefont {Freyse}},
  \bibinfo {author} {\bibfnamefont {A.~N.}\ \bibnamefont {Chaika}}, \bibinfo
  {author} {\bibfnamefont {C.}~\bibnamefont {Callaert}}, \bibinfo {author}
  {\bibfnamefont {A.~M.}\ \bibnamefont {Abakumov}}, \bibinfo {author}
  {\bibfnamefont {J.}~\bibnamefont {Hadermann}}, \bibinfo {author}
  {\bibfnamefont {A.}~\bibnamefont {Varykhalov}}, \bibinfo {author}
  {\bibfnamefont {E.~D.~L.}\ \bibnamefont {Rienks}}, \bibinfo {author}
  {\bibfnamefont {G.}~\bibnamefont {Bihlmayer}}, \bibinfo {author}
  {\bibfnamefont {S.}~\bibnamefont {Blügel}},\ and\ \bibinfo {author}
  {\bibfnamefont {O.}~\bibnamefont {Rader}},\ }\bibfield  {title} {\bibinfo
  {title} {Anomalous behavior of the electronic structure of
  {({Bi}$_{1-x}${In}$_x$)$_2${Se}$_3$} across the quantum phase transition from
  topological to trivial insulator},\ }\href
  {https://doi.org/10.1103/PhysRevB.98.235110} {\bibfield  {journal} {\bibinfo
  {journal} {Physical Review B}\ }\textbf {\bibinfo {volume} {98}},\ \bibinfo
  {pages} {235110} (\bibinfo {year} {2018})}\BibitemShut {NoStop}%
\bibitem [{\citenamefont {Berges}\ \emph {et~al.}(2019)\citenamefont {Berges},
  \citenamefont {van Loon}, \citenamefont {Schobert}, \citenamefont {Rösner},\
  and\ \citenamefont {Wehling}}]{2019BergesPhonon}%
  \BibitemOpen
  \bibfield  {author} {\bibinfo {author} {\bibfnamefont {J.}~\bibnamefont
  {Berges}}, \bibinfo {author} {\bibfnamefont {E.~G. C.~P.}\ \bibnamefont {van
  Loon}}, \bibinfo {author} {\bibfnamefont {A.}~\bibnamefont {Schobert}},
  \bibinfo {author} {\bibfnamefont {M.}~\bibnamefont {Rösner}},\ and\ \bibinfo
  {author} {\bibfnamefont {T.~O.}\ \bibnamefont {Wehling}},\ }\href@noop {}
  {\bibinfo {title} {Ab-initio phonon self-energies and fluctuation diagnostics
  of phonon anomalies: {Lattice} instabilities from {Dirac} pseudospin physics
  in transition-metal dichalcogenides}} (\bibinfo {year} {2019}),\ \Eprint
  {https://arxiv.org/abs/1911.02450} {arXiv:1911.02450 [cond-mat.str-el]}
  \BibitemShut {NoStop}%
\bibitem [{\citenamefont {Bahramy}\ \emph {et~al.}(2012)\citenamefont
  {Bahramy}, \citenamefont {Yang}, \citenamefont {Arita},\ and\ \citenamefont
  {Nagaosa}}]{2012Bahramy}%
  \BibitemOpen
  \bibfield  {author} {\bibinfo {author} {\bibfnamefont {M.}~\bibnamefont
  {Bahramy}}, \bibinfo {author} {\bibfnamefont {B.-J.}\ \bibnamefont {Yang}},
  \bibinfo {author} {\bibfnamefont {R.}~\bibnamefont {Arita}},\ and\ \bibinfo
  {author} {\bibfnamefont {N.}~\bibnamefont {Nagaosa}},\ }\bibfield  {title}
  {\bibinfo {title} {Emergence of non-centrosymmetric topological insulating
  phase in {BiTeI} under pressure},\ }\href
  {https://doi.org/10.1038/ncomms1679} {\bibfield  {journal} {\bibinfo
  {journal} {Nature Communications}\ }\textbf {\bibinfo {volume} {3}},\
  \bibinfo {pages} {679} (\bibinfo {year} {2012})}\BibitemShut {NoStop}%
\bibitem [{\citenamefont {Yang}\ \emph {et~al.}(2013)\citenamefont {Yang},
  \citenamefont {Bahramy}, \citenamefont {Arita}, \citenamefont {Isobe},
  \citenamefont {Moon},\ and\ \citenamefont {Nagaosa}}]{2013Yang}%
  \BibitemOpen
  \bibfield  {author} {\bibinfo {author} {\bibfnamefont {B.-J.}\ \bibnamefont
  {Yang}}, \bibinfo {author} {\bibfnamefont {M.~S.}\ \bibnamefont {Bahramy}},
  \bibinfo {author} {\bibfnamefont {R.}~\bibnamefont {Arita}}, \bibinfo
  {author} {\bibfnamefont {H.}~\bibnamefont {Isobe}}, \bibinfo {author}
  {\bibfnamefont {E.-G.}\ \bibnamefont {Moon}},\ and\ \bibinfo {author}
  {\bibfnamefont {N.}~\bibnamefont {Nagaosa}},\ }\bibfield  {title} {\bibinfo
  {title} {Theory of topological quantum phase transitions in {3D}
  noncentrosymmetric systems},\ }\href
  {https://doi.org/10.1103/PhysRevLett.110.086402} {\bibfield  {journal}
  {\bibinfo  {journal} {Physical Review Letters}\ }\textbf {\bibinfo {volume}
  {110}},\ \bibinfo {pages} {086402} (\bibinfo {year} {2013})}\BibitemShut
  {NoStop}%
\bibitem [{\citenamefont {Liu}\ and\ \citenamefont
  {Vanderbilt}(2014)}]{2014Liu}%
  \BibitemOpen
  \bibfield  {author} {\bibinfo {author} {\bibfnamefont {J.}~\bibnamefont
  {Liu}}\ and\ \bibinfo {author} {\bibfnamefont {D.}~\bibnamefont
  {Vanderbilt}},\ }\bibfield  {title} {\bibinfo {title} {{Weyl} semimetals from
  noncentrosymmetric topological insulators},\ }\href
  {https://doi.org/10.1103/PhysRevB.90.155316} {\bibfield  {journal} {\bibinfo
  {journal} {Physical Review B}\ }\textbf {\bibinfo {volume} {90}},\ \bibinfo
  {pages} {155316} (\bibinfo {year} {2014})}\BibitemShut {NoStop}%
\bibitem [{\citenamefont {Monserrat}\ and\ \citenamefont
  {Vanderbilt}(2017)}]{2017MonserratPRM}%
  \BibitemOpen
  \bibfield  {author} {\bibinfo {author} {\bibfnamefont {B.}~\bibnamefont
  {Monserrat}}\ and\ \bibinfo {author} {\bibfnamefont {D.}~\bibnamefont
  {Vanderbilt}},\ }\bibfield  {title} {\bibinfo {title} {Temperature dependence
  of the bulk {Rashba} splitting in the bismuth tellurohalides},\ }\href
  {https://doi.org/10.1103/PhysRevMaterials.1.054201} {\bibfield  {journal}
  {\bibinfo  {journal} {Physical Review Materials}\ }\textbf {\bibinfo {volume}
  {1}},\ \bibinfo {pages} {054201} (\bibinfo {year} {2017})}\BibitemShut
  {NoStop}%
\bibitem [{\citenamefont {Patrick}\ and\ \citenamefont
  {Giustino}(2013)}]{2013Patrick}%
  \BibitemOpen
  \bibfield  {author} {\bibinfo {author} {\bibfnamefont {C.~E.}\ \bibnamefont
  {Patrick}}\ and\ \bibinfo {author} {\bibfnamefont {F.}~\bibnamefont
  {Giustino}},\ }\bibfield  {title} {\bibinfo {title} {Quantum nuclear dynamics
  in the photophysics of diamondoids},\ }\href
  {https://doi.org/10.1038/ncomms3006} {\bibfield  {journal} {\bibinfo
  {journal} {Nature Communications}\ }\textbf {\bibinfo {volume} {4}},\
  \bibinfo {pages} {2006} (\bibinfo {year} {2013})}\BibitemShut {NoStop}%
\bibitem [{\citenamefont {Monserrat}(2016)}]{2016Monserrat_thermal_line}%
  \BibitemOpen
  \bibfield  {author} {\bibinfo {author} {\bibfnamefont {B.}~\bibnamefont
  {Monserrat}},\ }\bibfield  {title} {\bibinfo {title} {Vibrational averages
  along thermal lines},\ }\href {https://doi.org/10.1103/PhysRevB.93.014302}
  {\bibfield  {journal} {\bibinfo  {journal} {Physical Review B}\ }\textbf
  {\bibinfo {volume} {93}},\ \bibinfo {pages} {014302} (\bibinfo {year}
  {2016})}\BibitemShut {NoStop}%
\bibitem [{\citenamefont {Zacharias}\ \emph {et~al.}(2015)\citenamefont
  {Zacharias}, \citenamefont {Patrick},\ and\ \citenamefont
  {Giustino}}]{2015Zacharias}%
  \BibitemOpen
  \bibfield  {author} {\bibinfo {author} {\bibfnamefont {M.}~\bibnamefont
  {Zacharias}}, \bibinfo {author} {\bibfnamefont {C.~E.}\ \bibnamefont
  {Patrick}},\ and\ \bibinfo {author} {\bibfnamefont {F.}~\bibnamefont
  {Giustino}},\ }\bibfield  {title} {\bibinfo {title} {Stochastic approach to
  phonon-assisted optical absorption},\ }\href
  {https://doi.org/10.1103/PhysRevLett.115.177401} {\bibfield  {journal}
  {\bibinfo  {journal} {Physical Review Letters}\ }\textbf {\bibinfo {volume}
  {115}},\ \bibinfo {pages} {177401} (\bibinfo {year} {2015})}\BibitemShut
  {NoStop}%
\bibitem [{\citenamefont {Zacharias}\ and\ \citenamefont
  {Giustino}(2016)}]{2016Zacharias}%
  \BibitemOpen
  \bibfield  {author} {\bibinfo {author} {\bibfnamefont {M.}~\bibnamefont
  {Zacharias}}\ and\ \bibinfo {author} {\bibfnamefont {F.}~\bibnamefont
  {Giustino}},\ }\bibfield  {title} {\bibinfo {title} {One-shot calculation of
  temperature-dependent optical spectra and phonon-induced band-gap
  renormalization},\ }\href {https://doi.org/10.1103/PhysRevB.94.075125}
  {\bibfield  {journal} {\bibinfo  {journal} {Physical Review B}\ }\textbf
  {\bibinfo {volume} {94}},\ \bibinfo {pages} {075125} (\bibinfo {year}
  {2016})}\BibitemShut {NoStop}%
\bibitem [{\citenamefont {Zacharias}\ and\ \citenamefont
  {Giustino}(2019)}]{2019Zacharias}%
  \BibitemOpen
  \bibfield  {author} {\bibinfo {author} {\bibfnamefont {M.}~\bibnamefont
  {Zacharias}}\ and\ \bibinfo {author} {\bibfnamefont {F.}~\bibnamefont
  {Giustino}},\ }\bibfield  {title} {\bibinfo {title} {Theory of the special
  displacement method for electronic structure calculations at finite
  temperature},\ }\href {http://arxiv.org/abs/1912.10929} {\bibfield  {journal}
  {\bibinfo  {journal} {arXiv:1912.10929 [cond-mat]}\ } (\bibinfo {year}
  {2019})},\ \bibinfo {note} {arXiv: 1912.10929}\BibitemShut {NoStop}%
\bibitem [{\citenamefont {Della~Sala}\ \emph {et~al.}(2004)\citenamefont
  {Della~Sala}, \citenamefont {Rousseau}, \citenamefont {G\"orling},\ and\
  \citenamefont {Marx}}]{2004SalaPIMD}%
  \BibitemOpen
  \bibfield  {author} {\bibinfo {author} {\bibfnamefont {F.}~\bibnamefont
  {Della~Sala}}, \bibinfo {author} {\bibfnamefont {R.}~\bibnamefont
  {Rousseau}}, \bibinfo {author} {\bibfnamefont {A.}~\bibnamefont
  {G\"orling}},\ and\ \bibinfo {author} {\bibfnamefont {D.}~\bibnamefont
  {Marx}},\ }\bibfield  {title} {\bibinfo {title} {Quantum and thermal
  fluctuation effects on the photoabsorption spectra of clusters},\ }\href
  {https://doi.org/10.1103/PhysRevLett.92.183401} {\bibfield  {journal}
  {\bibinfo  {journal} {Phys. Rev. Lett.}\ }\textbf {\bibinfo {volume} {92}},\
  \bibinfo {pages} {183401} (\bibinfo {year} {2004})}\BibitemShut {NoStop}%
\bibitem [{\citenamefont {Ram\'{\i}rez}\ \emph {et~al.}(2006)\citenamefont
  {Ram\'{\i}rez}, \citenamefont {Herrero},\ and\ \citenamefont
  {Hern\'andez}}]{2006RamirezPIMD}%
  \BibitemOpen
  \bibfield  {author} {\bibinfo {author} {\bibfnamefont {R.}~\bibnamefont
  {Ram\'{\i}rez}}, \bibinfo {author} {\bibfnamefont {C.~P.}\ \bibnamefont
  {Herrero}},\ and\ \bibinfo {author} {\bibfnamefont {E.~R.}\ \bibnamefont
  {Hern\'andez}},\ }\bibfield  {title} {\bibinfo {title} {Path-integral
  molecular dynamics simulation of diamond},\ }\href
  {https://doi.org/10.1103/PhysRevB.73.245202} {\bibfield  {journal} {\bibinfo
  {journal} {Phys. Rev. B}\ }\textbf {\bibinfo {volume} {73}},\ \bibinfo
  {pages} {245202} (\bibinfo {year} {2006})}\BibitemShut {NoStop}%
\bibitem [{\citenamefont {Franceschetti}(2007)}]{2007FranceschettiMD}%
  \BibitemOpen
  \bibfield  {author} {\bibinfo {author} {\bibfnamefont {A.}~\bibnamefont
  {Franceschetti}},\ }\bibfield  {title} {\bibinfo {title} {First-principles
  calculations of the temperature dependence of the band gap of {Si}
  nanocrystals},\ }\href {https://doi.org/10.1103/PhysRevB.76.161301}
  {\bibfield  {journal} {\bibinfo  {journal} {Phys. Rev. B}\ }\textbf {\bibinfo
  {volume} {76}},\ \bibinfo {pages} {161301} (\bibinfo {year}
  {2007})}\BibitemShut {NoStop}%
\bibitem [{\citenamefont {Ram\'{\i}rez}\ \emph {et~al.}(2008)\citenamefont
  {Ram\'{\i}rez}, \citenamefont {Herrero}, \citenamefont {Hern\'andez},\ and\
  \citenamefont {Cardona}}]{2008RamirezPIMD}%
  \BibitemOpen
  \bibfield  {author} {\bibinfo {author} {\bibfnamefont {R.}~\bibnamefont
  {Ram\'{\i}rez}}, \bibinfo {author} {\bibfnamefont {C.~P.}\ \bibnamefont
  {Herrero}}, \bibinfo {author} {\bibfnamefont {E.~R.}\ \bibnamefont
  {Hern\'andez}},\ and\ \bibinfo {author} {\bibfnamefont {M.}~\bibnamefont
  {Cardona}},\ }\bibfield  {title} {\bibinfo {title} {Path-integral molecular
  dynamics simulation of $3c\text{\ensuremath{-}}\mathrm{Si}\mathrm{C}$},\
  }\href {https://doi.org/10.1103/PhysRevB.77.045210} {\bibfield  {journal}
  {\bibinfo  {journal} {Phys. Rev. B}\ }\textbf {\bibinfo {volume} {77}},\
  \bibinfo {pages} {045210} (\bibinfo {year} {2008})}\BibitemShut {NoStop}%
\bibitem [{\citenamefont {Gonze}(2020)}]{2020Gonze_totalenergy}%
  \BibitemOpen
  \bibfield  {author} {\bibinfo {author} {\bibfnamefont {X.}~\bibnamefont
  {Gonze}},\ }\bibfield  {title} {\bibinfo {title} {Predominance of
  non-adiabatic effects in zero-point renormalization of electronic energies},\
  }\href {http://totalenergy2020.dipc.org} {\bibfield  {journal} {\bibinfo
  {journal} {Workshop on Computational Physics and Materials Science, Total
  Energy and Force Methods}\ } (\bibinfo {year} {2020})}\BibitemShut {NoStop}%
\bibitem [{\citenamefont {Verdi}\ \emph {et~al.}(2017)\citenamefont {Verdi},
  \citenamefont {Caruso},\ and\ \citenamefont {Giustino}}]{2017VerdiPolaron}%
  \BibitemOpen
  \bibfield  {author} {\bibinfo {author} {\bibfnamefont {C.}~\bibnamefont
  {Verdi}}, \bibinfo {author} {\bibfnamefont {F.}~\bibnamefont {Caruso}},\ and\
  \bibinfo {author} {\bibfnamefont {F.}~\bibnamefont {Giustino}},\ }\bibfield
  {title} {\bibinfo {title} {Origin of the crossover from polarons to {Fermi}
  liquids in transition metal oxides},\ }\href
  {https://doi.org/10.1038/ncomms15769} {\bibfield  {journal} {\bibinfo
  {journal} {Nature Communications}\ }\textbf {\bibinfo {volume} {8}},\
  \bibinfo {pages} {15769} (\bibinfo {year} {2017})}\BibitemShut {NoStop}%
\bibitem [{Note1()}]{Note1}%
  \BibitemOpen
  \bibinfo {note} {See Supplemental Material, which includes Refs.~\cite
  {2009GiannozziQE, 2008Eiguren, 2019GonzeABINIT, 2011Gonze, 2013HamannONCVPSP,
  2018VanSettenPseudoDojo, 1996PerdewPBE, 2016Antonius, 2015PonceJCP,
  2015Verdi, 1997GonzePRB, 2006VanSchilfgaardeGW, 2018Nery, 2019BrownAltvater,
  1998ZhangrevPBE, 1976Allen, 2019QueralesFlores, 2017GiustinoRMP}, at [URL
  will be inserted by publisher] for the detailed derivation of Eq.~\protect
  \textup {\hbox {\mathsurround \z@ \protect \normalfont (\ignorespaces \ref
  {eq:dtilde_d}\unskip \@@italiccorr )}}, derivation of Eq.~\protect \textup
  {\hbox {\mathsurround \z@ \protect \normalfont (\ignorespaces \ref
  {eq:dtilde_d}\unskip \@@italiccorr )}} from Eqs.~\protect \textup {\hbox
  {\mathsurround \z@ \protect \normalfont (\ignorespaces \ref
  {eq:asr_1od}\unskip \@@italiccorr )}} and \protect \textup {\hbox
  {\mathsurround \z@ \protect \normalfont (\ignorespaces \ref
  {eq:dtilde_od}\unskip \@@italiccorr )}}, numerical tests of Eq.~\protect
  \textup {\hbox {\mathsurround \z@ \protect \normalfont (\ignorespaces \ref
  {eq:dtilde_od}\unskip \@@italiccorr )}}, the computational details, the
  convergence study of the electron self-energy, and the temperature dependence
  of the band gap of {\protect \text {BiTlSe$_2$}}.}\BibitemShut {Stop}%
\bibitem [{\citenamefont {Eremeev}\ \emph {et~al.}(2011)\citenamefont
  {Eremeev}, \citenamefont {Bihlmayer}, \citenamefont {Vergniory},
  \citenamefont {Koroteev}, \citenamefont {Menshchikova}, \citenamefont {Henk},
  \citenamefont {Ernst},\ and\ \citenamefont {Chulkov}}]{2011EremeevPRB}%
  \BibitemOpen
  \bibfield  {author} {\bibinfo {author} {\bibfnamefont {S.~V.}\ \bibnamefont
  {Eremeev}}, \bibinfo {author} {\bibfnamefont {G.}~\bibnamefont {Bihlmayer}},
  \bibinfo {author} {\bibfnamefont {M.}~\bibnamefont {Vergniory}}, \bibinfo
  {author} {\bibfnamefont {Y.~M.}\ \bibnamefont {Koroteev}}, \bibinfo {author}
  {\bibfnamefont {T.~V.}\ \bibnamefont {Menshchikova}}, \bibinfo {author}
  {\bibfnamefont {J.}~\bibnamefont {Henk}}, \bibinfo {author} {\bibfnamefont
  {A.}~\bibnamefont {Ernst}},\ and\ \bibinfo {author} {\bibfnamefont {E.~V.}\
  \bibnamefont {Chulkov}},\ }\bibfield  {title} {\bibinfo {title} {Ab initio
  electronic structure of thallium-based topological insulators},\ }\href
  {https://doi.org/10.1103/PhysRevB.83.205129} {\bibfield  {journal} {\bibinfo
  {journal} {Phys. Rev. B}\ }\textbf {\bibinfo {volume} {83}},\ \bibinfo
  {pages} {205129} (\bibinfo {year} {2011})}\BibitemShut {NoStop}%
\bibitem [{\citenamefont {Sato}\ \emph {et~al.}(2010)\citenamefont {Sato},
  \citenamefont {Segawa}, \citenamefont {Guo}, \citenamefont {Sugawara},
  \citenamefont {Souma}, \citenamefont {Takahashi},\ and\ \citenamefont
  {Ando}}]{2010SatoPRL}%
  \BibitemOpen
  \bibfield  {author} {\bibinfo {author} {\bibfnamefont {T.}~\bibnamefont
  {Sato}}, \bibinfo {author} {\bibfnamefont {K.}~\bibnamefont {Segawa}},
  \bibinfo {author} {\bibfnamefont {H.}~\bibnamefont {Guo}}, \bibinfo {author}
  {\bibfnamefont {K.}~\bibnamefont {Sugawara}}, \bibinfo {author}
  {\bibfnamefont {S.}~\bibnamefont {Souma}}, \bibinfo {author} {\bibfnamefont
  {T.}~\bibnamefont {Takahashi}},\ and\ \bibinfo {author} {\bibfnamefont
  {Y.}~\bibnamefont {Ando}},\ }\bibfield  {title} {\bibinfo {title} {Direct
  evidence for the {Dirac}-cone topological surface states in the ternary
  chalcogenide {${\mathrm{TlBiSe}}_{2}$}},\ }\href
  {https://doi.org/10.1103/PhysRevLett.105.136802} {\bibfield  {journal}
  {\bibinfo  {journal} {Phys. Rev. Lett.}\ }\textbf {\bibinfo {volume} {105}},\
  \bibinfo {pages} {136802} (\bibinfo {year} {2010})}\BibitemShut {NoStop}%
\bibitem [{\citenamefont {Kuroda}\ \emph {et~al.}(2010)\citenamefont {Kuroda},
  \citenamefont {Ye}, \citenamefont {Kimura}, \citenamefont {Eremeev},
  \citenamefont {Krasovskii}, \citenamefont {Chulkov}, \citenamefont {Ueda},
  \citenamefont {Miyamoto}, \citenamefont {Okuda}, \citenamefont {Shimada},
  \citenamefont {Namatame},\ and\ \citenamefont {Taniguchi}}]{2010KurodaPRL}%
  \BibitemOpen
  \bibfield  {author} {\bibinfo {author} {\bibfnamefont {K.}~\bibnamefont
  {Kuroda}}, \bibinfo {author} {\bibfnamefont {M.}~\bibnamefont {Ye}}, \bibinfo
  {author} {\bibfnamefont {A.}~\bibnamefont {Kimura}}, \bibinfo {author}
  {\bibfnamefont {S.~V.}\ \bibnamefont {Eremeev}}, \bibinfo {author}
  {\bibfnamefont {E.~E.}\ \bibnamefont {Krasovskii}}, \bibinfo {author}
  {\bibfnamefont {E.~V.}\ \bibnamefont {Chulkov}}, \bibinfo {author}
  {\bibfnamefont {Y.}~\bibnamefont {Ueda}}, \bibinfo {author} {\bibfnamefont
  {K.}~\bibnamefont {Miyamoto}}, \bibinfo {author} {\bibfnamefont
  {T.}~\bibnamefont {Okuda}}, \bibinfo {author} {\bibfnamefont
  {K.}~\bibnamefont {Shimada}}, \bibinfo {author} {\bibfnamefont
  {H.}~\bibnamefont {Namatame}},\ and\ \bibinfo {author} {\bibfnamefont
  {M.}~\bibnamefont {Taniguchi}},\ }\bibfield  {title} {\bibinfo {title}
  {Experimental realization of a three-dimensional topological insulator phase
  in ternary chalcogenide {${\mathrm{TlBiSe}}_{2}$}},\ }\href
  {https://doi.org/10.1103/PhysRevLett.105.146801} {\bibfield  {journal}
  {\bibinfo  {journal} {Phys. Rev. Lett.}\ }\textbf {\bibinfo {volume} {105}},\
  \bibinfo {pages} {146801} (\bibinfo {year} {2010})}\BibitemShut {NoStop}%
\bibitem [{\citenamefont {Chen}\ \emph {et~al.}(2010)\citenamefont {Chen},
  \citenamefont {Liu}, \citenamefont {Analytis}, \citenamefont {Chu},
  \citenamefont {Zhang}, \citenamefont {Yan}, \citenamefont {Mo}, \citenamefont
  {Moore}, \citenamefont {Lu}, \citenamefont {Fisher}, \citenamefont {Zhang},
  \citenamefont {Hussain},\ and\ \citenamefont {Shen}}]{2010ChenPRL}%
  \BibitemOpen
  \bibfield  {author} {\bibinfo {author} {\bibfnamefont {Y.~L.}\ \bibnamefont
  {Chen}}, \bibinfo {author} {\bibfnamefont {Z.~K.}\ \bibnamefont {Liu}},
  \bibinfo {author} {\bibfnamefont {J.~G.}\ \bibnamefont {Analytis}}, \bibinfo
  {author} {\bibfnamefont {J.-H.}\ \bibnamefont {Chu}}, \bibinfo {author}
  {\bibfnamefont {H.~J.}\ \bibnamefont {Zhang}}, \bibinfo {author}
  {\bibfnamefont {B.~H.}\ \bibnamefont {Yan}}, \bibinfo {author} {\bibfnamefont
  {S.-K.}\ \bibnamefont {Mo}}, \bibinfo {author} {\bibfnamefont {R.~G.}\
  \bibnamefont {Moore}}, \bibinfo {author} {\bibfnamefont {D.~H.}\ \bibnamefont
  {Lu}}, \bibinfo {author} {\bibfnamefont {I.~R.}\ \bibnamefont {Fisher}},
  \bibinfo {author} {\bibfnamefont {S.~C.}\ \bibnamefont {Zhang}}, \bibinfo
  {author} {\bibfnamefont {Z.}~\bibnamefont {Hussain}},\ and\ \bibinfo {author}
  {\bibfnamefont {Z.-X.}\ \bibnamefont {Shen}},\ }\bibfield  {title} {\bibinfo
  {title} {Single {Dirac} cone topological surface state and unusual
  thermoelectric property of compounds from a new topological insulator
  family},\ }\href {https://doi.org/10.1103/PhysRevLett.105.266401} {\bibfield
  {journal} {\bibinfo  {journal} {Phys. Rev. Lett.}\ }\textbf {\bibinfo
  {volume} {105}},\ \bibinfo {pages} {266401} (\bibinfo {year}
  {2010})}\BibitemShut {NoStop}%
\bibitem [{Note5()}]{Note5}%
  \BibitemOpen
  \bibinfo {note} {The custom code developed in this work may be made available
  in a later release of the {\protect \sc Quantum ESPRESSO}\ package after
  discussions with the {\protect \sc Quantum ESPRESSO}\ core
  developers.}\BibitemShut {Stop}%
\bibitem [{\citenamefont {Fu}\ and\ \citenamefont {Kane}(2007)}]{2007Fu}%
  \BibitemOpen
  \bibfield  {author} {\bibinfo {author} {\bibfnamefont {L.}~\bibnamefont
  {Fu}}\ and\ \bibinfo {author} {\bibfnamefont {C.~L.}\ \bibnamefont {Kane}},\
  }\bibfield  {title} {\bibinfo {title} {Topological insulators with inversion
  symmetry},\ }\href {https://doi.org/10.1103/PhysRevB.76.045302} {\bibfield
  {journal} {\bibinfo  {journal} {Phys. Rev. B}\ }\textbf {\bibinfo {volume}
  {76}},\ \bibinfo {pages} {045302} (\bibinfo {year} {2007})}\BibitemShut
  {NoStop}%
\bibitem [{\citenamefont {Zhang}\ \emph {et~al.}(2014)\citenamefont {Zhang},
  \citenamefont {Liu}, \citenamefont {Luo}, \citenamefont {Freeman},\ and\
  \citenamefont {Zunger}}]{2014ZhangHSPth}%
  \BibitemOpen
  \bibfield  {author} {\bibinfo {author} {\bibfnamefont {X.}~\bibnamefont
  {Zhang}}, \bibinfo {author} {\bibfnamefont {Q.}~\bibnamefont {Liu}}, \bibinfo
  {author} {\bibfnamefont {J.-W.}\ \bibnamefont {Luo}}, \bibinfo {author}
  {\bibfnamefont {A.~J.}\ \bibnamefont {Freeman}},\ and\ \bibinfo {author}
  {\bibfnamefont {A.}~\bibnamefont {Zunger}},\ }\bibfield  {title} {\bibinfo
  {title} {Hidden spin polarization in inversion-symmetric bulk crystals},\
  }\href {https://doi.org/10.1038/nphys2933} {\bibfield  {journal} {\bibinfo
  {journal} {Nature Physics}\ }\textbf {\bibinfo {volume} {10}},\ \bibinfo
  {pages} {387} (\bibinfo {year} {2014})}\BibitemShut {NoStop}%
\bibitem [{\citenamefont {Riley}\ \emph {et~al.}(2014)\citenamefont {Riley},
  \citenamefont {Mazzola}, \citenamefont {Dendzik}, \citenamefont {Michiardi},
  \citenamefont {Takayama}, \citenamefont {Bawden}, \citenamefont {Granerød},
  \citenamefont {Leandersson}, \citenamefont {Balasubramanian}, \citenamefont
  {Hoesch}, \citenamefont {Kim}, \citenamefont {Takagi}, \citenamefont
  {Meevasana}, \citenamefont {Hofmann}, \citenamefont {Bahramy}, \citenamefont
  {Wells},\ and\ \citenamefont {King}}]{2014RileyHSPexp}%
  \BibitemOpen
  \bibfield  {author} {\bibinfo {author} {\bibfnamefont {J.~M.}\ \bibnamefont
  {Riley}}, \bibinfo {author} {\bibfnamefont {F.}~\bibnamefont {Mazzola}},
  \bibinfo {author} {\bibfnamefont {M.}~\bibnamefont {Dendzik}}, \bibinfo
  {author} {\bibfnamefont {M.}~\bibnamefont {Michiardi}}, \bibinfo {author}
  {\bibfnamefont {T.}~\bibnamefont {Takayama}}, \bibinfo {author}
  {\bibfnamefont {L.}~\bibnamefont {Bawden}}, \bibinfo {author} {\bibfnamefont
  {C.}~\bibnamefont {Granerød}}, \bibinfo {author} {\bibfnamefont
  {M.}~\bibnamefont {Leandersson}}, \bibinfo {author} {\bibfnamefont
  {T.}~\bibnamefont {Balasubramanian}}, \bibinfo {author} {\bibfnamefont
  {M.}~\bibnamefont {Hoesch}}, \bibinfo {author} {\bibfnamefont {T.~K.}\
  \bibnamefont {Kim}}, \bibinfo {author} {\bibfnamefont {H.}~\bibnamefont
  {Takagi}}, \bibinfo {author} {\bibfnamefont {W.}~\bibnamefont {Meevasana}},
  \bibinfo {author} {\bibfnamefont {P.}~\bibnamefont {Hofmann}}, \bibinfo
  {author} {\bibfnamefont {M.}~\bibnamefont {Bahramy}}, \bibinfo {author}
  {\bibfnamefont {J.}~\bibnamefont {Wells}},\ and\ \bibinfo {author}
  {\bibfnamefont {P.~C.}\ \bibnamefont {King}},\ }\bibfield  {title} {\bibinfo
  {title} {Direct observation of spin-polarized bulk bands in an
  inversion-symmetric semiconductor},\ }\href
  {https://doi.org/10.1038/nphys3105} {\bibfield  {journal} {\bibinfo
  {journal} {Nature Physics}\ }\textbf {\bibinfo {volume} {10}},\ \bibinfo
  {pages} {835} (\bibinfo {year} {2014})}\BibitemShut {NoStop}%
\bibitem [{\citenamefont {Yao}\ \emph {et~al.}(2017)\citenamefont {Yao},
  \citenamefont {Wang}, \citenamefont {Huang}, \citenamefont {Deng},
  \citenamefont {Yan}, \citenamefont {Zhang}, \citenamefont {Miyamoto},
  \citenamefont {Okuda}, \citenamefont {Li}, \citenamefont {Wang} \emph
  {et~al.}}]{2017YaoHSPOL}%
  \BibitemOpen
  \bibfield  {author} {\bibinfo {author} {\bibfnamefont {W.}~\bibnamefont
  {Yao}}, \bibinfo {author} {\bibfnamefont {E.}~\bibnamefont {Wang}}, \bibinfo
  {author} {\bibfnamefont {H.}~\bibnamefont {Huang}}, \bibinfo {author}
  {\bibfnamefont {K.}~\bibnamefont {Deng}}, \bibinfo {author} {\bibfnamefont
  {M.}~\bibnamefont {Yan}}, \bibinfo {author} {\bibfnamefont {K.}~\bibnamefont
  {Zhang}}, \bibinfo {author} {\bibfnamefont {K.}~\bibnamefont {Miyamoto}},
  \bibinfo {author} {\bibfnamefont {T.}~\bibnamefont {Okuda}}, \bibinfo
  {author} {\bibfnamefont {L.}~\bibnamefont {Li}}, \bibinfo {author}
  {\bibfnamefont {Y.}~\bibnamefont {Wang}}, \emph {et~al.},\ }\bibfield
  {title} {\bibinfo {title} {Direct observation of spin-layer locking by local
  {Rashba} effect in monolayer semiconducting {PtSe$_2$} film},\ }\href
  {https://doi.org/10.1038/ncomms14216} {\bibfield  {journal} {\bibinfo
  {journal} {Nature communications}\ }\textbf {\bibinfo {volume} {8}},\
  \bibinfo {pages} {14216} (\bibinfo {year} {2017})}\BibitemShut {NoStop}%
\bibitem [{\citenamefont {Razzoli}\ \emph {et~al.}(2017)\citenamefont
  {Razzoli}, \citenamefont {Jaouen}, \citenamefont {Mottas}, \citenamefont
  {Hildebrand}, \citenamefont {Monney}, \citenamefont {Pisoni}, \citenamefont
  {Muff}, \citenamefont {Fanciulli}, \citenamefont {Plumb}, \citenamefont
  {Rogalev}, \citenamefont {Strocov}, \citenamefont {Mesot}, \citenamefont
  {Shi}, \citenamefont {Dil}, \citenamefont {Beck},\ and\ \citenamefont
  {Aebi}}]{2017RazzoliHSPOL}%
  \BibitemOpen
  \bibfield  {author} {\bibinfo {author} {\bibfnamefont {E.}~\bibnamefont
  {Razzoli}}, \bibinfo {author} {\bibfnamefont {T.}~\bibnamefont {Jaouen}},
  \bibinfo {author} {\bibfnamefont {M.-L.}\ \bibnamefont {Mottas}}, \bibinfo
  {author} {\bibfnamefont {B.}~\bibnamefont {Hildebrand}}, \bibinfo {author}
  {\bibfnamefont {G.}~\bibnamefont {Monney}}, \bibinfo {author} {\bibfnamefont
  {A.}~\bibnamefont {Pisoni}}, \bibinfo {author} {\bibfnamefont
  {S.}~\bibnamefont {Muff}}, \bibinfo {author} {\bibfnamefont {M.}~\bibnamefont
  {Fanciulli}}, \bibinfo {author} {\bibfnamefont {N.~C.}\ \bibnamefont
  {Plumb}}, \bibinfo {author} {\bibfnamefont {V.~A.}\ \bibnamefont {Rogalev}},
  \bibinfo {author} {\bibfnamefont {V.~N.}\ \bibnamefont {Strocov}}, \bibinfo
  {author} {\bibfnamefont {J.}~\bibnamefont {Mesot}}, \bibinfo {author}
  {\bibfnamefont {M.}~\bibnamefont {Shi}}, \bibinfo {author} {\bibfnamefont
  {J.~H.}\ \bibnamefont {Dil}}, \bibinfo {author} {\bibfnamefont
  {H.}~\bibnamefont {Beck}},\ and\ \bibinfo {author} {\bibfnamefont
  {P.}~\bibnamefont {Aebi}},\ }\bibfield  {title} {\bibinfo {title} {Selective
  probing of hidden spin-polarized states in inversion-symmetric bulk
  {${\mathrm{MoS}}_{2}$}},\ }\href
  {https://doi.org/10.1103/PhysRevLett.118.086402} {\bibfield  {journal}
  {\bibinfo  {journal} {Phys. Rev. Lett.}\ }\textbf {\bibinfo {volume} {118}},\
  \bibinfo {pages} {086402} (\bibinfo {year} {2017})}\BibitemShut {NoStop}%
\bibitem [{\citenamefont {Cheng}\ \emph {et~al.}(2018)\citenamefont {Cheng},
  \citenamefont {Sun}, \citenamefont {Chen},\ and\ \citenamefont
  {Meng}}]{2018ChengHSPOL}%
  \BibitemOpen
  \bibfield  {author} {\bibinfo {author} {\bibfnamefont {C.}~\bibnamefont
  {Cheng}}, \bibinfo {author} {\bibfnamefont {J.-T.}\ \bibnamefont {Sun}},
  \bibinfo {author} {\bibfnamefont {X.-R.}\ \bibnamefont {Chen}},\ and\
  \bibinfo {author} {\bibfnamefont {S.}~\bibnamefont {Meng}},\ }\bibfield
  {title} {\bibinfo {title} {Hidden spin polarization in the {1$T$}-phase
  layered transition-metal dichalcogenides {MX$_2$ (M = Zr, Hf; X = S, Se,
  Te)}},\ }\href {https://doi.org/https://doi.org/10.1016/j.scib.2017.12.003}
  {\bibfield  {journal} {\bibinfo  {journal} {Science Bulletin}\ }\textbf
  {\bibinfo {volume} {63}},\ \bibinfo {pages} {85 } (\bibinfo {year}
  {2018})}\BibitemShut {NoStop}%
\bibitem [{\citenamefont {Oliva}\ \emph {et~al.}(2020)\citenamefont {Oliva},
  \citenamefont {Wo\'zniak}, \citenamefont {Dybala}, \citenamefont {Kopaczek},
  \citenamefont {Scharoch},\ and\ \citenamefont {Kudrawiec}}]{2020OlivaHSPOL}%
  \BibitemOpen
  \bibfield  {author} {\bibinfo {author} {\bibfnamefont {R.}~\bibnamefont
  {Oliva}}, \bibinfo {author} {\bibfnamefont {T.}~\bibnamefont {Wo\'zniak}},
  \bibinfo {author} {\bibfnamefont {F.}~\bibnamefont {Dybala}}, \bibinfo
  {author} {\bibfnamefont {J.}~\bibnamefont {Kopaczek}}, \bibinfo {author}
  {\bibfnamefont {P.}~\bibnamefont {Scharoch}},\ and\ \bibinfo {author}
  {\bibfnamefont {R.}~\bibnamefont {Kudrawiec}},\ }\bibfield  {title} {\bibinfo
  {title} {Hidden spin-polarized bands in semiconducting {2$H$}-{MoTe$_2$}},\
  }\href {https://doi.org/10.1080/21663831.2019.1702113} {\bibfield  {journal}
  {\bibinfo  {journal} {Materials Research Letters}\ }\textbf {\bibinfo
  {volume} {8}},\ \bibinfo {pages} {75} (\bibinfo {year} {2020})}\BibitemShut
  {NoStop}%
\bibitem [{\citenamefont {Tu}\ \emph {et~al.}(2020)\citenamefont {Tu},
  \citenamefont {Chen}, \citenamefont {Ruan}, \citenamefont {Zhao},
  \citenamefont {Xu}, \citenamefont {Chen}, \citenamefont {Zhang},
  \citenamefont {Zhang}, \citenamefont {Wu}, \citenamefont {He}, \citenamefont
  {Zhang}, \citenamefont {Zhang},\ and\ \citenamefont {Xu}}]{2020TuHSPOL}%
  \BibitemOpen
  \bibfield  {author} {\bibinfo {author} {\bibfnamefont {J.}~\bibnamefont
  {Tu}}, \bibinfo {author} {\bibfnamefont {X.~B.}\ \bibnamefont {Chen}},
  \bibinfo {author} {\bibfnamefont {X.~Z.}\ \bibnamefont {Ruan}}, \bibinfo
  {author} {\bibfnamefont {Y.~F.}\ \bibnamefont {Zhao}}, \bibinfo {author}
  {\bibfnamefont {H.~F.}\ \bibnamefont {Xu}}, \bibinfo {author} {\bibfnamefont
  {Z.~D.}\ \bibnamefont {Chen}}, \bibinfo {author} {\bibfnamefont {X.~Q.}\
  \bibnamefont {Zhang}}, \bibinfo {author} {\bibfnamefont {X.~W.}\ \bibnamefont
  {Zhang}}, \bibinfo {author} {\bibfnamefont {J.}~\bibnamefont {Wu}}, \bibinfo
  {author} {\bibfnamefont {L.}~\bibnamefont {He}}, \bibinfo {author}
  {\bibfnamefont {Y.}~\bibnamefont {Zhang}}, \bibinfo {author} {\bibfnamefont
  {R.}~\bibnamefont {Zhang}},\ and\ \bibinfo {author} {\bibfnamefont {Y.~B.}\
  \bibnamefont {Xu}},\ }\bibfield  {title} {\bibinfo {title} {Direct
  observation of hidden spin polarization in
  {$2H\text{\ensuremath{-}}\mathrm{MoT}{\mathrm{e}}_{2}$}},\ }\href
  {https://doi.org/10.1103/PhysRevB.101.035102} {\bibfield  {journal} {\bibinfo
   {journal} {Phys. Rev. B}\ }\textbf {\bibinfo {volume} {101}},\ \bibinfo
  {pages} {035102} (\bibinfo {year} {2020})}\BibitemShut {NoStop}%
\bibitem [{\citenamefont {Soluyanov}\ and\ \citenamefont
  {Vanderbilt}(2011)}]{Soluyanov2011Z2}%
  \BibitemOpen
  \bibfield  {author} {\bibinfo {author} {\bibfnamefont {A.~A.}\ \bibnamefont
  {Soluyanov}}\ and\ \bibinfo {author} {\bibfnamefont {D.}~\bibnamefont
  {Vanderbilt}},\ }\bibfield  {title} {\bibinfo {title} {Computing topological
  invariants without inversion symmetry},\ }\href
  {https://doi.org/10.1103/PhysRevB.83.235401} {\bibfield  {journal} {\bibinfo
  {journal} {Phys. Rev. B}\ }\textbf {\bibinfo {volume} {83}},\ \bibinfo
  {pages} {235401} (\bibinfo {year} {2011})}\BibitemShut {NoStop}%
\bibitem [{\citenamefont {Yu}\ \emph {et~al.}(2011)\citenamefont {Yu},
  \citenamefont {Qi}, \citenamefont {Bernevig}, \citenamefont {Fang},\ and\
  \citenamefont {Dai}}]{Yu2011Z2}%
  \BibitemOpen
  \bibfield  {author} {\bibinfo {author} {\bibfnamefont {R.}~\bibnamefont
  {Yu}}, \bibinfo {author} {\bibfnamefont {X.~L.}\ \bibnamefont {Qi}}, \bibinfo
  {author} {\bibfnamefont {A.}~\bibnamefont {Bernevig}}, \bibinfo {author}
  {\bibfnamefont {Z.}~\bibnamefont {Fang}},\ and\ \bibinfo {author}
  {\bibfnamefont {X.}~\bibnamefont {Dai}},\ }\bibfield  {title} {\bibinfo
  {title} {Equivalent expression of ${\mathbb{z}}_{2}$ topological invariant
  for band insulators using the non-abelian berry connection},\ }\href
  {https://doi.org/10.1103/PhysRevB.84.075119} {\bibfield  {journal} {\bibinfo
  {journal} {Phys. Rev. B}\ }\textbf {\bibinfo {volume} {84}},\ \bibinfo
  {pages} {075119} (\bibinfo {year} {2011})}\BibitemShut {NoStop}%
\bibitem [{\citenamefont {Ringel}\ and\ \citenamefont
  {Kraus}(2011)}]{Ringel2011Z2}%
  \BibitemOpen
  \bibfield  {author} {\bibinfo {author} {\bibfnamefont {Z.}~\bibnamefont
  {Ringel}}\ and\ \bibinfo {author} {\bibfnamefont {Y.~E.}\ \bibnamefont
  {Kraus}},\ }\bibfield  {title} {\bibinfo {title} {Determining topological
  order from a local ground-state correlation function},\ }\href
  {https://doi.org/10.1103/PhysRevB.83.245115} {\bibfield  {journal} {\bibinfo
  {journal} {Phys. Rev. B}\ }\textbf {\bibinfo {volume} {83}},\ \bibinfo
  {pages} {245115} (\bibinfo {year} {2011})}\BibitemShut {NoStop}%
\bibitem [{\citenamefont {Fukui}\ and\ \citenamefont
  {Hatsugai}(2007)}]{Fukui2007Z2}%
  \BibitemOpen
  \bibfield  {author} {\bibinfo {author} {\bibfnamefont {T.}~\bibnamefont
  {Fukui}}\ and\ \bibinfo {author} {\bibfnamefont {Y.}~\bibnamefont
  {Hatsugai}},\ }\bibfield  {title} {\bibinfo {title} {Quantum spin hall effect
  in three dimensional materials: Lattice computation of z2 topological
  invariants and its application to bi and sb},\ }\href
  {https://doi.org/https://doi.org/10.1143/jpsj.76.053702} {\bibfield
  {journal} {\bibinfo  {journal} {Journal of the Physical Society of Japan}\
  }\textbf {\bibinfo {volume} {76}},\ \bibinfo {pages} {053702} (\bibinfo
  {year} {2007})}\BibitemShut {NoStop}%
\bibitem [{\citenamefont {Giannozzi}\ \emph {et~al.}(2009)\citenamefont
  {Giannozzi}, \citenamefont {Baroni}, \citenamefont {Bonini}, \citenamefont
  {Calandra}, \citenamefont {Car}, \citenamefont {Cavazzoni}, \citenamefont
  {Ceresoli}, \citenamefont {Chiarotti}, \citenamefont {Cococcioni},
  \citenamefont {Dabo}, \citenamefont {Dal~Corso}, \citenamefont
  {de~Gironcoli}, \citenamefont {Fabris}, \citenamefont {Fratesi},
  \citenamefont {Gebauer}, \citenamefont {Gerstmann}, \citenamefont
  {Gougoussis}, \citenamefont {Kokalj}, \citenamefont {Lazzeri}, \citenamefont
  {Martin-Samos}, \citenamefont {Marzari}, \citenamefont {Mauri}, \citenamefont
  {Mazzarello}, \citenamefont {Paolini}, \citenamefont {Pasquarello},
  \citenamefont {Paulatto}, \citenamefont {Sbraccia}, \citenamefont {Scandolo},
  \citenamefont {Sclauzero}, \citenamefont {Seitsonen}, \citenamefont
  {Smogunov}, \citenamefont {Umari},\ and\ \citenamefont
  {Wentzcovitch}}]{2009GiannozziQE}%
  \BibitemOpen
  \bibfield  {author} {\bibinfo {author} {\bibfnamefont {P.}~\bibnamefont
  {Giannozzi}}, \bibinfo {author} {\bibfnamefont {S.}~\bibnamefont {Baroni}},
  \bibinfo {author} {\bibfnamefont {N.}~\bibnamefont {Bonini}}, \bibinfo
  {author} {\bibfnamefont {M.}~\bibnamefont {Calandra}}, \bibinfo {author}
  {\bibfnamefont {R.}~\bibnamefont {Car}}, \bibinfo {author} {\bibfnamefont
  {C.}~\bibnamefont {Cavazzoni}}, \bibinfo {author} {\bibfnamefont
  {D.}~\bibnamefont {Ceresoli}}, \bibinfo {author} {\bibfnamefont {G.~L.}\
  \bibnamefont {Chiarotti}}, \bibinfo {author} {\bibfnamefont {M.}~\bibnamefont
  {Cococcioni}}, \bibinfo {author} {\bibfnamefont {I.}~\bibnamefont {Dabo}},
  \bibinfo {author} {\bibfnamefont {A.}~\bibnamefont {Dal~Corso}}, \bibinfo
  {author} {\bibfnamefont {S.}~\bibnamefont {de~Gironcoli}}, \bibinfo {author}
  {\bibfnamefont {S.}~\bibnamefont {Fabris}}, \bibinfo {author} {\bibfnamefont
  {G.}~\bibnamefont {Fratesi}}, \bibinfo {author} {\bibfnamefont
  {R.}~\bibnamefont {Gebauer}}, \bibinfo {author} {\bibfnamefont
  {U.}~\bibnamefont {Gerstmann}}, \bibinfo {author} {\bibfnamefont
  {C.}~\bibnamefont {Gougoussis}}, \bibinfo {author} {\bibfnamefont
  {A.}~\bibnamefont {Kokalj}}, \bibinfo {author} {\bibfnamefont
  {M.}~\bibnamefont {Lazzeri}}, \bibinfo {author} {\bibfnamefont
  {L.}~\bibnamefont {Martin-Samos}}, \bibinfo {author} {\bibfnamefont
  {N.}~\bibnamefont {Marzari}}, \bibinfo {author} {\bibfnamefont
  {F.}~\bibnamefont {Mauri}}, \bibinfo {author} {\bibfnamefont
  {R.}~\bibnamefont {Mazzarello}}, \bibinfo {author} {\bibfnamefont
  {S.}~\bibnamefont {Paolini}}, \bibinfo {author} {\bibfnamefont
  {A.}~\bibnamefont {Pasquarello}}, \bibinfo {author} {\bibfnamefont
  {L.}~\bibnamefont {Paulatto}}, \bibinfo {author} {\bibfnamefont
  {C.}~\bibnamefont {Sbraccia}}, \bibinfo {author} {\bibfnamefont
  {S.}~\bibnamefont {Scandolo}}, \bibinfo {author} {\bibfnamefont
  {G.}~\bibnamefont {Sclauzero}}, \bibinfo {author} {\bibfnamefont {A.~P.}\
  \bibnamefont {Seitsonen}}, \bibinfo {author} {\bibfnamefont {A.}~\bibnamefont
  {Smogunov}}, \bibinfo {author} {\bibfnamefont {P.}~\bibnamefont {Umari}},\
  and\ \bibinfo {author} {\bibfnamefont {R.}~\bibnamefont {Wentzcovitch}},\
  }\bibfield  {title} {\bibinfo {title} {{QUANTUM ESPRESSO: a modular and
  open-source software project for quantum simulations of materials}},\ }\href
  {http://stacks.iop.org/0953-8984/21/i=39/a=395502} {\bibfield  {journal}
  {\bibinfo  {journal} {J. Phys.: Condens. Matter}\ }\textbf {\bibinfo {volume}
  {21}},\ \bibinfo {pages} {395502} (\bibinfo {year} {2009})}\BibitemShut
  {NoStop}%
\bibitem [{\citenamefont {Eiguren}\ and\ \citenamefont
  {Ambrosch-Draxl}(2008)}]{2008Eiguren}%
  \BibitemOpen
  \bibfield  {author} {\bibinfo {author} {\bibfnamefont {A.}~\bibnamefont
  {Eiguren}}\ and\ \bibinfo {author} {\bibfnamefont {C.}~\bibnamefont
  {Ambrosch-Draxl}},\ }\bibfield  {title} {\bibinfo {title} {Wannier
  interpolation scheme for phonon-induced potentials: Application to bulk
  {${\text{MgB}}_{2}$}, {W}, and the $(1\ifmmode\times\else\texttimes\fi{}1)$
  {H}-covered {W}(110) surface},\ }\href
  {https://doi.org/10.1103/PhysRevB.78.045124} {\bibfield  {journal} {\bibinfo
  {journal} {Phys. Rev. B}\ }\textbf {\bibinfo {volume} {78}},\ \bibinfo
  {pages} {045124} (\bibinfo {year} {2008})}\BibitemShut {NoStop}%
\bibitem [{\citenamefont {Gonze}\ \emph {et~al.}(2019)\citenamefont {Gonze},
  \citenamefont {Amadon}, \citenamefont {Antonius}, \citenamefont {Arnardi},
  \citenamefont {Baguet}, \citenamefont {Beuken}, \citenamefont {Bieder},
  \citenamefont {Bottin}, \citenamefont {Bouchet}, \citenamefont {Bousquet}
  \emph {et~al.}}]{2019GonzeABINIT}%
  \BibitemOpen
  \bibfield  {author} {\bibinfo {author} {\bibfnamefont {X.}~\bibnamefont
  {Gonze}}, \bibinfo {author} {\bibfnamefont {B.}~\bibnamefont {Amadon}},
  \bibinfo {author} {\bibfnamefont {G.}~\bibnamefont {Antonius}}, \bibinfo
  {author} {\bibfnamefont {F.}~\bibnamefont {Arnardi}}, \bibinfo {author}
  {\bibfnamefont {L.}~\bibnamefont {Baguet}}, \bibinfo {author} {\bibfnamefont
  {J.-M.}\ \bibnamefont {Beuken}}, \bibinfo {author} {\bibfnamefont
  {J.}~\bibnamefont {Bieder}}, \bibinfo {author} {\bibfnamefont
  {F.}~\bibnamefont {Bottin}}, \bibinfo {author} {\bibfnamefont
  {J.}~\bibnamefont {Bouchet}}, \bibinfo {author} {\bibfnamefont
  {E.}~\bibnamefont {Bousquet}}, \emph {et~al.},\ }\bibfield  {title} {\bibinfo
  {title} {The {Abinit} project: Impact, environment and recent developments},\
  }\href {https://doi.org/10.1016/j.cpc.2019.107042} {\bibfield  {journal}
  {\bibinfo  {journal} {Computer Physics Communications}\ ,\ \bibinfo {pages}
  {107042}} (\bibinfo {year} {2019})}\BibitemShut {NoStop}%
\bibitem [{\citenamefont {Hamann}(2013)}]{2013HamannONCVPSP}%
  \BibitemOpen
  \bibfield  {author} {\bibinfo {author} {\bibfnamefont {D.~R.}\ \bibnamefont
  {Hamann}},\ }\bibfield  {title} {\bibinfo {title} {Optimized norm-conserving
  {Vanderbilt} pseudopotentials},\ }\href
  {https://doi.org/10.1103/PhysRevB.88.085117} {\bibfield  {journal} {\bibinfo
  {journal} {Physical Review B}\ }\textbf {\bibinfo {volume} {88}},\ \bibinfo
  {pages} {085117} (\bibinfo {year} {2013})}\BibitemShut {NoStop}%
\bibitem [{\citenamefont {van Setten}\ \emph {et~al.}(2018)\citenamefont {van
  Setten}, \citenamefont {Giantomassi}, \citenamefont {Bousquet}, \citenamefont
  {Verstraete}, \citenamefont {Hamann}, \citenamefont {Gonze},\ and\
  \citenamefont {Rignanese}}]{2018VanSettenPseudoDojo}%
  \BibitemOpen
  \bibfield  {author} {\bibinfo {author} {\bibfnamefont {M.}~\bibnamefont {van
  Setten}}, \bibinfo {author} {\bibfnamefont {M.}~\bibnamefont {Giantomassi}},
  \bibinfo {author} {\bibfnamefont {E.}~\bibnamefont {Bousquet}}, \bibinfo
  {author} {\bibfnamefont {M.}~\bibnamefont {Verstraete}}, \bibinfo {author}
  {\bibfnamefont {D.}~\bibnamefont {Hamann}}, \bibinfo {author} {\bibfnamefont
  {X.}~\bibnamefont {Gonze}},\ and\ \bibinfo {author} {\bibfnamefont {G.-M.}\
  \bibnamefont {Rignanese}},\ }\bibfield  {title} {\bibinfo {title} {The
  {PseudoDojo}: Training and grading a 85 element optimized norm-conserving
  pseudopotential table},\ }\href {https://doi.org/10.1016/j.cpc.2018.01.012}
  {\bibfield  {journal} {\bibinfo  {journal} {Computer Physics Communications}\
  }\textbf {\bibinfo {volume} {226}},\ \bibinfo {pages} {39} (\bibinfo {year}
  {2018})}\BibitemShut {NoStop}%
\bibitem [{\citenamefont {Perdew}\ \emph {et~al.}(1996)\citenamefont {Perdew},
  \citenamefont {Burke},\ and\ \citenamefont {Ernzerhof}}]{1996PerdewPBE}%
  \BibitemOpen
  \bibfield  {author} {\bibinfo {author} {\bibfnamefont {J.~P.}\ \bibnamefont
  {Perdew}}, \bibinfo {author} {\bibfnamefont {K.}~\bibnamefont {Burke}},\ and\
  \bibinfo {author} {\bibfnamefont {M.}~\bibnamefont {Ernzerhof}},\ }\bibfield
  {title} {\bibinfo {title} {Generalized gradient approximation made simple},\
  }\href {https://doi.org/10.1103/PhysRevLett.77.3865} {\bibfield  {journal}
  {\bibinfo  {journal} {Phys. Rev. Lett.}\ }\textbf {\bibinfo {volume} {77}},\
  \bibinfo {pages} {3865} (\bibinfo {year} {1996})}\BibitemShut {NoStop}%
\bibitem [{\citenamefont {Verdi}\ and\ \citenamefont
  {Giustino}(2015)}]{2015Verdi}%
  \BibitemOpen
  \bibfield  {author} {\bibinfo {author} {\bibfnamefont {C.}~\bibnamefont
  {Verdi}}\ and\ \bibinfo {author} {\bibfnamefont {F.}~\bibnamefont
  {Giustino}},\ }\bibfield  {title} {\bibinfo {title} {Fr\"ohlich
  electron-phonon vertex from first principles},\ }\href
  {https://doi.org/10.1103/PhysRevLett.115.176401} {\bibfield  {journal}
  {\bibinfo  {journal} {Phys. Rev. Lett.}\ }\textbf {\bibinfo {volume} {115}},\
  \bibinfo {pages} {176401} (\bibinfo {year} {2015})}\BibitemShut {NoStop}%
\bibitem [{\citenamefont {Gonze}(1997)}]{1997GonzePRB}%
  \BibitemOpen
  \bibfield  {author} {\bibinfo {author} {\bibfnamefont {X.}~\bibnamefont
  {Gonze}},\ }\bibfield  {title} {\bibinfo {title} {First-principles responses
  of solids to atomic displacements and homogeneous electric fields:
  Implementation of a conjugate-gradient algorithm},\ }\href
  {https://doi.org/10.1103/PhysRevB.55.10337} {\bibfield  {journal} {\bibinfo
  {journal} {Phys. Rev. B}\ }\textbf {\bibinfo {volume} {55}},\ \bibinfo
  {pages} {10337} (\bibinfo {year} {1997})}\BibitemShut {NoStop}%
\bibitem [{\citenamefont {van Schilfgaarde}\ \emph {et~al.}(2006)\citenamefont
  {van Schilfgaarde}, \citenamefont {Kotani},\ and\ \citenamefont
  {Faleev}}]{2006VanSchilfgaardeGW}%
  \BibitemOpen
  \bibfield  {author} {\bibinfo {author} {\bibfnamefont {M.}~\bibnamefont {van
  Schilfgaarde}}, \bibinfo {author} {\bibfnamefont {T.}~\bibnamefont
  {Kotani}},\ and\ \bibinfo {author} {\bibfnamefont {S.}~\bibnamefont
  {Faleev}},\ }\bibfield  {title} {\bibinfo {title} {Quasiparticle
  self-consistent {$GW$} theory},\ }\href
  {https://doi.org/10.1103/PhysRevLett.96.226402} {\bibfield  {journal}
  {\bibinfo  {journal} {Phys. Rev. Lett.}\ }\textbf {\bibinfo {volume} {96}},\
  \bibinfo {pages} {226402} (\bibinfo {year} {2006})}\BibitemShut {NoStop}%
\bibitem [{\citenamefont {Brown-Altvater}(2019)}]{2019BrownAltvater}%
  \BibitemOpen
  \bibfield  {author} {\bibinfo {author} {\bibfnamefont {F.}~\bibnamefont
  {Brown-Altvater}},\ }\emph {\bibinfo {title} {Electronic excitations,
  phonons, and electron-phonon coupling in acenes}},\ \href@noop {} {Ph.D.
  thesis},\ \bibinfo  {school} {University of California, Berkeley} (\bibinfo
  {year} {2019})\BibitemShut {NoStop}%
\bibitem [{\citenamefont {Zhang}\ and\ \citenamefont
  {Yang}(1998)}]{1998ZhangrevPBE}%
  \BibitemOpen
  \bibfield  {author} {\bibinfo {author} {\bibfnamefont {Y.}~\bibnamefont
  {Zhang}}\ and\ \bibinfo {author} {\bibfnamefont {W.}~\bibnamefont {Yang}},\
  }\bibfield  {title} {\bibinfo {title} {Comment on ``generalized gradient
  approximation made simple''},\ }\href
  {https://doi.org/10.1103/PhysRevLett.80.890} {\bibfield  {journal} {\bibinfo
  {journal} {Phys. Rev. Lett.}\ }\textbf {\bibinfo {volume} {80}},\ \bibinfo
  {pages} {890} (\bibinfo {year} {1998})}\BibitemShut {NoStop}%
\end{thebibliography}%


%apsrev4-2.bst 2019-01-14 (MD) hand-edited version of apsrev4-1.bst
%Control: key (0)
%Control: author (8) initials jnrlst
%Control: editor formatted (1) identically to author
%Control: production of article title (0) allowed
%Control: page (0) single
%Control: year (1) truncated
%Control: production of eprint (0) enabled
%

\end{document}

\title{Supplemental Material: Theory of phonon-induced renormalization of electron wavefunctions}

\author{Jae-Mo Lihm}
\author{Cheol-Hwan Park}
\email{cheolhwan@snu.ac.kr}
\affiliation{Department of Physics and Astronomy, Seoul National University, Seoul 08826, Korea}
\affiliation{Center for Correlated Electron Systems, Institute for Basic Science, Seoul 08826, Korea}
\affiliation{Center for Theoretical Physics, Seoul National University, Seoul 08826, Korea}

\date{\today}

\maketitle

%%%%%%%%%%%%%%%%%%%%%%%%%%%%%%%%%%%%%%%%%%%%%%%%%%%%%%%%%%%%%%%%%%%%%%%%%
\section{Crystal structure and Brillouin zone of \NoCaseChange{\bts}.}

\begin{figure}[!htbp]
\includegraphics[width=0.5\columnwidth]{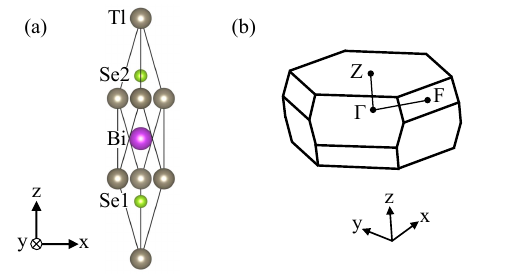}
\caption{
(a) Crystal structure of \bts. (b) Brillouin zone of \bts.
}
\label{fig:bts}
\end{figure}

%%%%%%%%%%%%%%%%%%%%%%%%%%%%%%%%%%%%%%%%%%%%%%%%%%%%%%%%%%%%%%%%%%%%%%%%%
\section{A detailed derivation of Eq.~(\ref{eq:dtilde_d})}
Let us consider the displaced system discussed in the main text, above Eq.~\eqref{eq:veps_tinv}.
From second-order perturbation theory assuming nondegenerate states, the eigenvalues of the displaced system read
\begin{align} \label{eq:S_veps_series}
    &\veps^{\kappa\alpha;\alpha'}_\nk(\xi, \tau)
    = \veps_\nk^{(0)} + \xi \mel{u_{\nk}}{\duGka \vks}{u_\nk} + \tau \sum_{\kappa'} \mel{u_{\nk}}{\duGkpap \vks}{u_\nk} \\
    &+ \xi \tau \sum_{\kappa'}\left( \mel{u_{\nk}}{\duGka \duGkpap \vks}{u_{\nk}}
    + 2 \Re \sum_{m\neq n} 
    \frac{\mel{u_\nk}{\duGka \vks}{u_\mk} \mel{u_\mk}{\duGkpap \vks}{u_\nk}}{\veps_\nk^{(0)} - \veps_\mk^{(0)}} \right)
    + \mathcal{O}(\xi^2) + \mathcal{O}(\tau^2), \nonumber
\end{align}
with $\veps_\nk^{(0)} = \veps^{\kappa\alpha;\alpha'}_\nk(\xi=0,\tau=0)$.

Now, the translational invariance of the eigenenergies [Eq.~\eqref{eq:veps_tinv}] implies that all coefficients of the terms including $\tau$ in Eq.~\eqref{eq:S_veps_series} should vanish.
By applying this condition to the coefficients of $\tau$, one finds
\begin{equation} \label{eq:S_asr_1d}
    \sum_{\kappa'} \mel{u_{\nk}}{\duGkpap \vks}{u_\nk} = 0.
\end{equation}
Similarly, from the coefficients of $\xi\tau$, one finds
\begin{equation} \label{eq:S_asr_2d}
    \sum_{\kappa'} \mel{u_{\nk}}{\duGka \duGkpap \vks}{u_{\nk}}
    =- 2 \Re \sum_{m\neq n} 
    \frac{\mel{u_\nk}{\duGka \vks}{u_\mk} \mel{u_\mk}{\duGkpap \vks}{u_\nk}}{\veps_\nk^{(0)} - \veps_\mk^{(0)}}.
\end{equation}
Equations~\eqref{eq:S_asr_1d} and \eqref{eq:S_asr_2d} are the first- and second-order acoustic sum rules for the diagonal EPC~\cite{2017GiustinoRMP}.

By substituting the left-hand side of Eq.~\eqref{eq:S_asr_2d} to the second line of Eq.~\eqref{eq:d_ria_def}, one finds Eq.~\eqref{eq:dtilde_d}, an expression for the diagonal part of $\dtilde$ that does not refer to the second-order derivative of the KS potential.

%%%%%%%%%%%%%%%%%%%%%%%%%%%%%%%%%%%%%%%%%%%%%%%%%%%%%%%%%%%%%%%%%%%%%%%%%
\section{An alternative derivation of Eq.~(\ref{eq:dtilde_d}) using our results}
In this section, we provide an alternative derivation of Eq.~\eqref{eq:dtilde_d} using the main results of our theory [Eqs.~\eqref{eq:asr_1od}, \eqref{eq:asr_2od} and \eqref{eq:dtilde_od}].

First, by expanding the right hand side of Eq.~\eqref{eq:dtilde_od} in the electron energy eigenbasis, one finds
\begin{align} \label{eq:S_dtilde_nn}
    \dtilde^{\kappa\alpha\alpha'}_{nn'}(\mb{k})
    &= i \mel{u_\nk}{[\duGka \vks, \hat{p}_{\alpha'}]}{u_{\npk}} \nnnl
    &= i \sum_{m} \left( \mel{u_\nk}{\duGka \vks}{u_\mk} \mel{u_\mk}{\hat{p}_{\alpha'}}{u_\npk}
    - \mel{u_\nk}{\hat{p}_{\alpha'}}{u_\mk} \mel{u_\mk}{\duGka \vks}{u_\npk} \right) \nnnl
    &= i \sum_{m} \left[ h_{nm}^{\kappa\alpha}(\mb{k},\mb{\Gamma}) \mel{u_\mk}{\hat{p}_{\alpha'}}{u_\npk}
    - \mel{u_\nk}{\hat{p}_{\alpha'}}{u_\mk} h_{mn'}^{\kappa\alpha}(\mb{k},\mb{\Gamma}) \right].
\end{align}
To evaluate $\mel{u_\nk}{\hat{p}_{\alpha'}}{u_\mk}$, we use Eq.~\eqref{eq:asr_1od}, the first-order acoustic sum rule in the operator form.
Taking the matrix element of Eq.~\eqref{eq:asr_1od}, one finds
\begin{equation} \label{eq:S_asr_1od_mel}
    \sum_{\kappa'} h_{nm}^{\kappa'\alpha'}(\mb{k},\mb{\Gamma})
    = i \mel{u_\nk}{[\vks, \hat{p}_{\alpha'}]}{u_\mk}
    = i \mel{u_\nk}{[\hat{H}, \hat{p}_{\alpha'}]}{u_\mk}
    = i (\veps_\nk - \veps_\mk) \mel{u_\nk}{\hat{p}_{\alpha'}}{u_\mk}.
\end{equation}
Here, we used the fact that the electron Hamiltonian $\hat{H}$ is the sum of the KS potential $\vks$ and the kinetic energy $\hat{T}=\hat{\mb{p}}^2/2m$ and that the kinetic energy operator commutes with $\hat{p}_{\alpha'}$.

Rearranging Eq.~\eqref{eq:S_asr_1od_mel}, one finds that if $\veps_\mk \neq \veps_\nk$, the matrix element of the momentum operator becomes
\begin{equation} \label{eq:S_p_mel}
    \mel{u_\nk}{\hat{p}_{\alpha'}}{u_\mk}
    = -i \sum_{\kappa'} \frac{h_{nm}^{\kappa'\alpha'}(\mb{k},\mb{\Gamma})}{\veps_\nk - \veps_\mk}.
\end{equation}
Note that one cannot use Eq.~\eqref{eq:S_p_mel} if $m=n$.
Also, we assume that the electron states are nondegenerate.
The formalism can be extended to degenerate cases by first choosing a different wavefunction gauge for each $\alpha'$ so that the off-diagonal matrix elements of $\hat{p}_{\alpha'}$ between degenerate states vanish and then applying unitary rotations to the matrix elements $\dtilde^{\kappa\alpha\alpha'}_{nn'}(\mb{k})$ so that the wavefunction gauge for different $\alpha'$ becomes identical.

By substituting Eq.~\eqref{eq:S_p_mel} into Eq.~\eqref{eq:S_dtilde_nn}, one finds
\begin{align} \label{eq:S_dtilde_nn'}
    \dtilde^{\kappa\alpha\alpha'}_{nn'}(\mb{k})
    =& \sum_{\kappa'}\Bigg[
    h_{nn'}^{\kappa\alpha}(\mb{k},\mb{\Gamma}) \Big( \mel{u_\npk}{\hat{p}_{\alpha'}}{u_\npk}
    - \mel{u_\nk}{\hat{p}_{\alpha'}}{u_\nk} \Big) \nnnl
    &+ \sum_{m \neq n'}  \frac{h_{nm}^{\kappa\alpha}(\mb{k},\mb{\Gamma}) h_{mn'}^{\kappa'\alpha'}(\mb{k},\mb{\Gamma})}{\veps_\mk - \veps_{\npk}}
    - \sum_{m \neq n}  \frac{h_{nm}^{\kappa'\alpha'}(\mb{k},\mb{\Gamma}) h_{mn'}^{\kappa\alpha}(\mb{k},\mb{\Gamma})}{\veps_\nk - \veps_\mk}
    \Bigg].
\end{align}

For the diagonal case of $n=n'$, the first term in the parenthesis of Eq.~\eqref{eq:S_dtilde_nn'} cancels out.
Thus, the diagonal case of Eq.~\eqref{eq:S_dtilde_nn'} is equivalent to Eq.~\eqref{eq:dtilde_d}, the expression of the original AHC theory for the diagonal DW matrix element.
In contrast, the first term of Eq.~\eqref{eq:S_dtilde_nn'} does not cancel out for the off-diagonal case, $n\neq n'$.
Thus, the expression of the original AHC theory [Eq.~\eqref{eq:dtilde_d}] cannot be generalized to the off-diagonal case without using our operator-generalized formulation.

%%%%%%%%%%%%%%%%%%%%%%%%%%%%%%%%%%%%%%%%%%%%%%%%%%%%%%%%%%%%%%%%%%%%%%%%%
\section{Numerical tests of Eq.~(\ref{eq:dtilde_od})} \label{sec:S_test}
\begin{figure}[h]
\includegraphics[width=0.5\columnwidth]{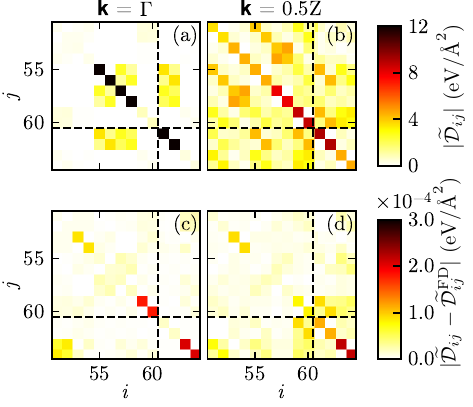}
\caption{
(a-b) Absolute values and (c-d) absolute errors of the second-order EPC matrix elements $\dtilde$ of \bts{} computed using Eq.~\eqref{eq:dtilde_od} and the finite-difference (FD) method.
The dashed black lines indicate the boundary between the highest valence band and the lowest conduction band. The second-order matrix element $\dtilde$ is computed for the mixed derivative for the displacement of the Tl atoms along the $x$ direction and the uniform displacement of all atoms along the $x$ direction.
}
\label{fig:S_bts_test}
\end{figure}

\begin{figure}[h]
\includegraphics[width=0.5\columnwidth]{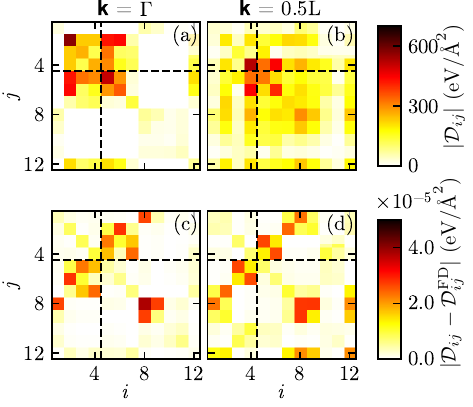}
\caption{
(a-b) Absolute values and (c-d) absolute errors of the second-order EPC matrix elements $\dtilde$ of diamond computed using Eq.~\eqref{eq:dtilde_od} and the finite-difference method.
The dashed black lines indicate the boundary between the highest valence band and the lowest conduction band.
The second-order matrix element $\dtilde$ is computed for the mixed derivative for the displacement of all lattice-periodic images of one C atom along the $x$ direction and the uniform displacement of all atoms along the $x$ direction.
}
\label{fig:S_diamond}
\end{figure}

To numerically test the theory developed in this work, we compute the second-order matrix elements $\dtilde$ using Eq.~\eqref{eq:dtilde_od} as well as the finite-difference method.
The finite-difference calculations were converged with a Richardson interpolation of order 4.
The results of the comparison for \bts{} and diamond are shown in Fig.~\ref{fig:S_bts_test} and Fig.~\ref{fig:S_diamond}, respectively.
Both results show an excellent agreement between Eq.~\eqref{eq:dtilde_od} and the finite-difference calculations, with errors several orders of magnitude smaller than the typical value of $\dtilde$.
This result provides a numerical confirmation of our theory.

%%%%%%%%%%%%%%%%%%%%%%%%%%%%%%%%%%%%%%%%%%%%%%%%%%%%%%%%%%%%%%%%%%%%%%%%%
\section{Computational Details}
We used the \qe\ package~\cite{2009GiannozziQE} to perform DFT and DFPT calculations.
We modified the \qe\ package to perform the Fourier interpolation of phonon potential~\cite{2008Eiguren,2019GonzeABINIT} and to circumvent the summation over a large number of unoccupied states in the Fan self-energy by solving the Sternheimer equation~\cite{2011Gonze}.
The custom code developed in this work may be made available in a later release of the \qe\ package after discussions with the \qe\ core developers.

The DFT and DFPT calculations were performed with an 8$\times$8$\times$8 $k$-point grid, a kinetic energy cutoff of 80~Ry, 
fully relativistic ONCV pseudopotentials~\cite{2013HamannONCVPSP} taken from the PseudoDojo library (v0.4)~\cite{2018VanSettenPseudoDojo}, and the PBE functional~\cite{1996PerdewPBE}.
The optimized lattice parameters were taken from Ref.~\cite{2016Antonius}.
In calculating the temperature dependence, we neglected the effect of thermal expansion as done in Ref.~\cite{2016Antonius}.

To avoid a sum over a large number of high-energy empty bands in the Fan self-energy, we approximated the contribution of the high-energy bands using the solution of the Sternheimer equation~\cite{2011Gonze}.
We call the contribution of the high-energy bands to the Fan self-energy the ``upper Fan'' self-energy and the contribution from the low-energy bands the ``lower Fan'' self-energy.
For the lower Fan self-energy, we included an artificial broadening $\eta$ to smooth the energy denominators.
The phonon frequency in the denominator of Eq.~\eqref{eq:fan_def} was ignored in the calculation of the upper Fan self-energy.
We included 30 unoccupied bands in the calculation of the lower Fan self-energy to include all states lying less than 10~eV above the valence band maximum energy.
The resulting error due to the static approximation of the upper Fan self-energy is at most a few meV.
We note that our expression for the DW self-energy [Eq.~\eqref{eq:dtilde_od}] does not require a sum over bands nor solutions of the Sternheimer equations, contrary to the ordinary expressions~\cite{2011Gonze,2015PonceJCP}.

The potential perturbation is computed on a coarse $6 \times 6 \times 6$ $q$-point grid from DFPT and then Fourier interpolated~\cite{2008Eiguren} to the fine $q$-point grid.
The nonanalytic long-range part of the potential perturbation is taken into account by subtracting and adding the dipole potential~\cite{2015Verdi} before and after the Fourier interpolation, respectively, following the method of Ref.~\cite{2019GonzeABINIT}.
The force constants are Fourier interpolated to the fine $q$-point grid using the dynamical matrices computed at the $6 \times 6 \times 6$ grid.
We take the non-analytic long-range part of the force constants correctly into account~\cite{1997GonzePRB}.
We used a $8\times8\times8$ $q$-point grid for the calculation of the upper Fan and DW self-energies, and a $36\times36\times36$ $q$-point grid for the calculation of the lower Fan self-energy.
We also included the broadening of $\eta=20~{\rm meV}$ in the calculation of the lower Fan self-energy. 
The sizes of the fine $q$-point grids and the broadening parameter $\eta$ were determined after a thorough convergence study (see Sec.~\ref{sec:S_conv}).

We calculated the self-energy matrix in the basis of 10 valence and 4 conduction bands to include all states within 2\,eV from the valence band maximum energy
We have also checked that the inclusion of only two valence and two conduction bands is already sufficient to give converged renormalized electron energies.

The self-energy matrix we compute is not diagonal and not Hermitian. Thus, the corresponding eigenvectors are not orthonormal.
Also, since the self-energy is dynamical, the choice of the frequency for the off-diagonal components is ambiguous.
Following the recipe established by the {\it GW} community~\cite{2006VanSchilfgaardeGW}, we use a static Hermitian approximation of the self-energy:
\begin{equation} \label{eq:sigma_herm}
    \Sigma^{\rm Herm.}_{nn'\mb{k}}
    = \frac{1}{2} \left[ (\Re \Sigma)_{nn'\mb{k}}(\veps_{\nk}) + (\Re \Sigma)_{nn'\mb{k}}(\veps_{\npk}) \right].
\end{equation}
Here, $\Re \Sigma$ is defined as follows:
\begin{equation} \label{eq:sigma_re}
    (\Re \Sigma)_{nn'\mb{k}} = \frac{1}{2} \left( \Sigma_{\mb{k}} + \Sigma_{\mb{k}}^\dagger \right)_{nn'}.
\end{equation}
In the $GW$ literature, $\Sigma^{\rm Herm.}$ defined in Eq.~\eqref{eq:sigma_herm} is known to be a nearly optimal approximant of the full dynamical self-energy~\cite{2006VanSchilfgaardeGW}.
We then diagonalize the renormalized Hamiltonian
\begin{equation} \label{eq:e_plus_sigma}
    H^{\rm renorm.}_{nn'\mb{k}} = \veps_{\nk}\delta_{n,n'} + \Sigma^{\rm Herm.}_{nn'\mb{k}}
\end{equation}
to obtain the renormalized electron energies.
This method is a generalization of the on-the-mass-shell approximation~\cite{2018Nery}
to the case of an off-diagonal self-energy.

An alternate method to deal with a dynamical self-energy is the Dyson-Migdal approach that self-consistently solves the eigenvalue equation
\begin{equation}
    \det \left[ \Sigma_{\mb{k}}(\omega=E_\nk) - E_\nk \mathbb{1} \right] = 0
\end{equation}
with $\mathbb{1}$ the identity matrix.
However, it has been found that the on-the-mass-shell approximation gives a much better estimate of renormalized quasiparticle energies than the Dyson-Migdal approach~\cite{2018Nery,2019BrownAltvater}.
Thus, we use the on-the-mass-shell approximation in this work.
In passing we note that our method of calculating the off-diagonal matrix elements of the self-energy is universal and can be applied to other many-body theories than the static Hermitian approximation [Eq.~\eqref{eq:sigma_herm}] and the on-the-mass-shell approximation [Eq.~\eqref{eq:e_plus_sigma}] we adopted here. Although important, the comparison between the results obtained from different levels of many-body theories is not the main subject of our paper. Our paper provides a method that enables such future investigations.

%%%%%%%%%%%%%%%%%%%%%%%%%%%%%%%%%%%%%%%%%%%%%%%%%%%%%%%%%%%%%%%%%%%%%%%%%
\section{Convergence study of the electron self-energy} \label{sec:S_conv}
\begin{figure}
\includegraphics[width=0.9\columnwidth]{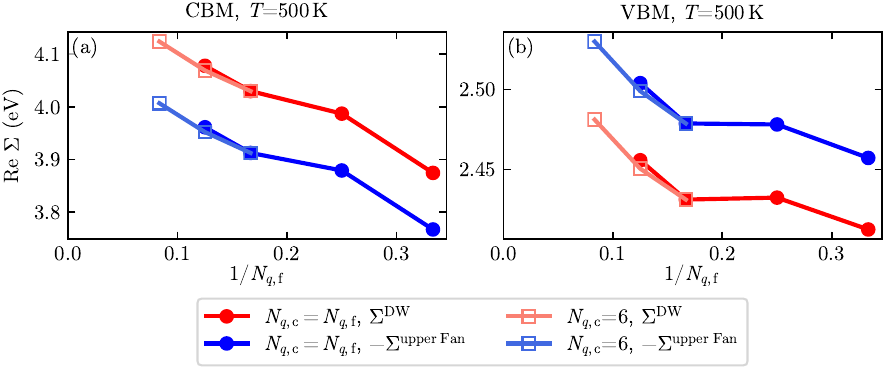}
\caption{
Convergence study for the DW and upper Fan self-energies at the (a) CBM and (b) VBM at 500~K.
}
\label{fig:S_conv_each}
\end{figure}
\begin{figure}
\includegraphics[width=0.9\columnwidth]{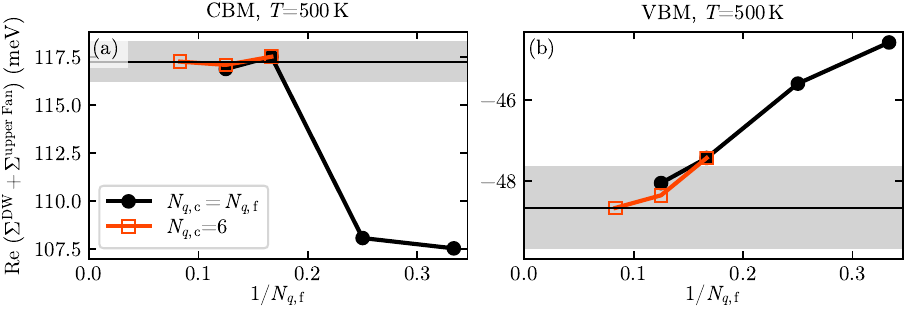}
\caption{
Convergence study for the sum of the DW and upper Fan self-energies at the (a) CBM and (b) VBM at 500~K.
The horizontal solid line indicates the self-energy computed with $N_{q,\rm{c}}=6$ and $N_{q,\rm{f}}=12$.
The shaded area indicates the range where the absolute error with the value indicated by the horizontal solid line is below 1~meV.
}
\label{fig:S_conv_dw_fan}
\end{figure}
\begin{figure*}
\includegraphics[width=0.8\textwidth]{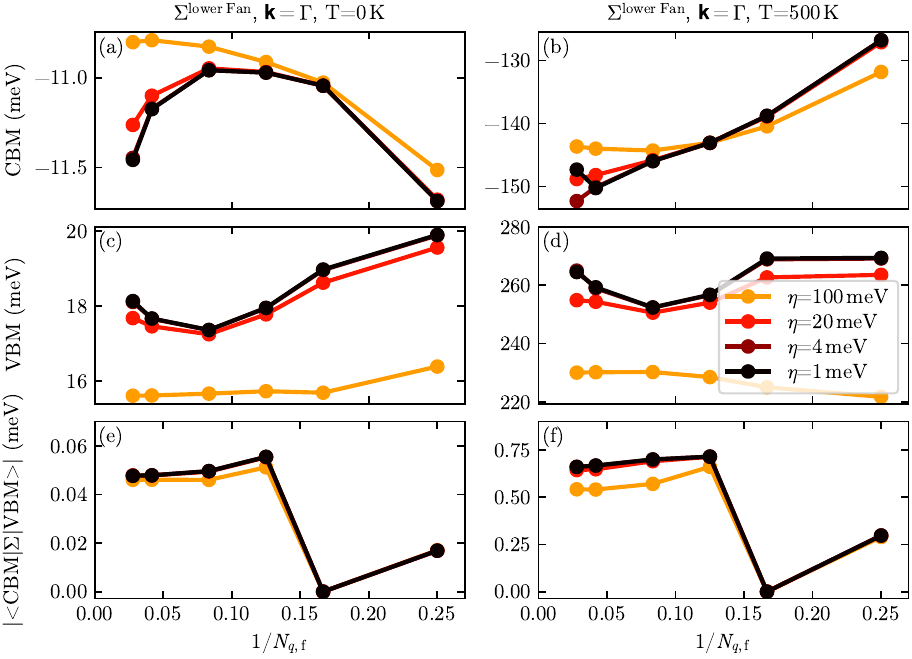}
\caption{
Convergence study for the lower Fan self-energy of the (a-b) CBM, (c-d) VBM, and (e-f) off-diagonal element at $\mb{k}={\rm \Gamma}$ at (a, c, e) 0~K and (b, d, f) 500~K.
The real part of the self-energy is plotted for the case of the CBM and VBM, and the absolute value of the self-energy is plotted for the off-diagonal element.
The maximal value of $\Nqf$ shown is $\Nqf=36$.
}
\label{fig:S_fan_gamma}
\end{figure*}
\begin{figure*}
\includegraphics[width=0.8\textwidth]{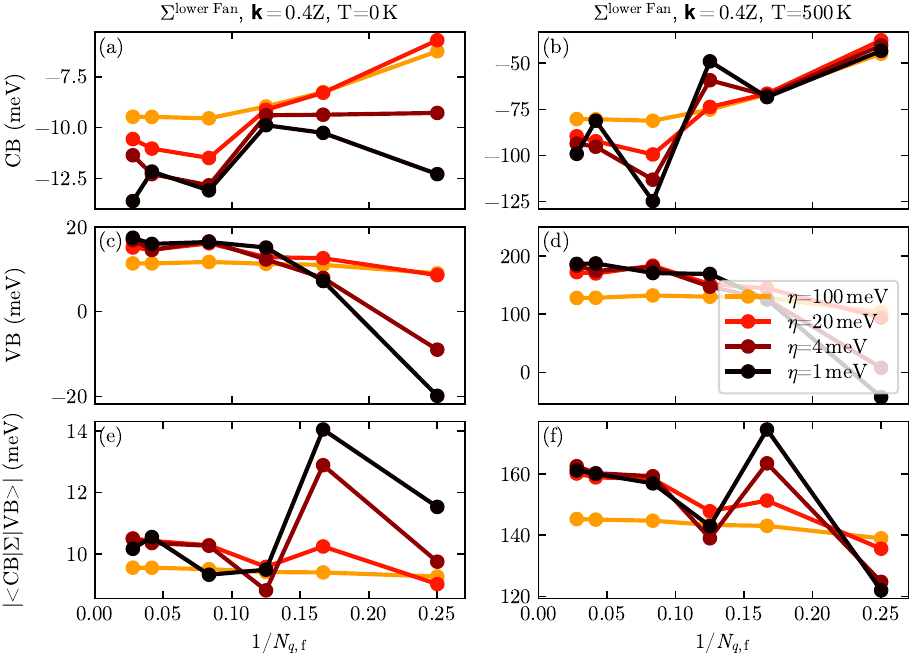}
\caption{
Convergence study for the lower Fan self-energy of the (a-b) lowest conduction band (CB), (c-d) highest valence band (VB), and (e-f) off-diagonal coupling between the CB and VB at $\mb{k}=0.4{\rm Z}=(0.0, 0.0, 0.16)\,{\rm \AA^{-1}}$ at (a, c, e) 0~K and (b, d, f) 500~K.
The real part of the self-energy is plotted for the case of the CB and VB, and the absolute value of the self-energy is plotted for the off-diagonal part.
The maximal value of $\Nqf$ shown is $\Nqf=36$.
}
\label{fig:S_fan_04z}
\end{figure*}

In this section, we describe the convergence study for the sampling of phonon wave vectors.
To study the convergence of the electron self-energy, we computed the potential perturbation on a coarse $\Nqc \times \Nqc \times \Nqc$ $q$-point grid from DFPT and then Fourier interpolated it to a fine $\Nqf \times \Nqf \times \Nqf$ $q$-point grid for different values of $\Nqc$ and $\Nqf$.

Figures~\ref{fig:S_conv_each} and \ref{fig:S_conv_dw_fan} show the convergence of the DW and upper Fan self-energies and their sum.
We find that the convergence of both the DW and upper Fan self-energies themselves [Fig.~\ref{fig:S_conv_each}] is much slower than that of the sum of these two terms [Fig.~\ref{fig:S_conv_dw_fan}].
Note that the vertical scales of Fig.~\ref{fig:S_conv_dw_fan} is more than an order of magnitude smaller than those of Fig.~\ref{fig:S_conv_each}.
This observation can be partly attributed to the cancellation between the DW self-energy and the Fan self-energy in the long-wavelength limit~\cite{1976Allen,2011Gonze,2019QueralesFlores}.
Thus, it suffices to converge the sum of the DW and upper Fan self-energies, not their individual contributions.

The sum of the DW and upper Fan self-energies is converged at $\Nqf=8$ with an error less than 1~meV.
We also find that the self-energy computed using the phonon potential interpolated from $\Nqc=6$ to $\Nqf=8$ gives reasonable agreement with the self-energy computed with $\Nqc=\Nqf=8$ without any interpolation.
We checked that every matrix element between the highest valence band and the lowest conduction band of the sum of the DW and upper Fan self-energies is converged for all temperatures below 500~K and for all $k$ points plotted in Fig.~\ref{fig:band} with an error less than 2~meV.
From these analyses, we choose to use $\Nqc=6$ and $\Nqf=8$ for the evaluation of the sum of the DW and upper Fan self-energies.

Figures~\ref{fig:S_fan_gamma} and \ref{fig:S_fan_04z} show the convergence of the lower Fan self-energy with respect to the size of the fine grid and the artificial broadening $\eta$.
We used $\Nqc=6$ for all cases.
Comparing the scales of the left columns [(a,c,e)] with the scales of the right columns [(b,d,f)] of Fig.~\ref{fig:S_fan_gamma} and \ref{fig:S_fan_04z}, we find that the absolute error coming from an insufficient convergence is much larger at $T=500~{\rm K}$ than at $T=0~{\rm K}$.
Also, by comparing Fig.~\ref{fig:S_fan_gamma}(b) and Fig.~\ref{fig:S_fan_04z}(b), we find that the self-energy of the lowest conduction band at $k =0.4{\rm Z}$ is much more oscillatory and slowly convergent with respect to $\Nqf$ than the self-energy at $k={\rm \Gamma}$.
We believe that the slow convergence can be attributed to the existence of near-resonant transitions from the lowest conduction band at $k=0.4{\rm Z}$ to other conduction band states, which is possible as the state at $k=0.4{\rm Z}$ is not a band extremum.
At $\Nqf=36$, which is the largest $q$-point grid size we have used, we find that an artificial broadening of at least 20~meV is required to obtain a smooth band structure.
Therefore, we choose to use $\Nqf=36$ and $\eta=$20~meV for the lower Fan self-energy as a compromise of accuracy and computational cost.

%%%%%%%%%%%%%%%%%%%%%%%%%%%%%%%%%%%%%%%%%%%%%%%%%%%%%%%%%%%%%%%%%%%%%%%%%
\section{Temperature dependence of the band gap}
\begin{figure}[!htbp]
\includegraphics[width=0.5\columnwidth]{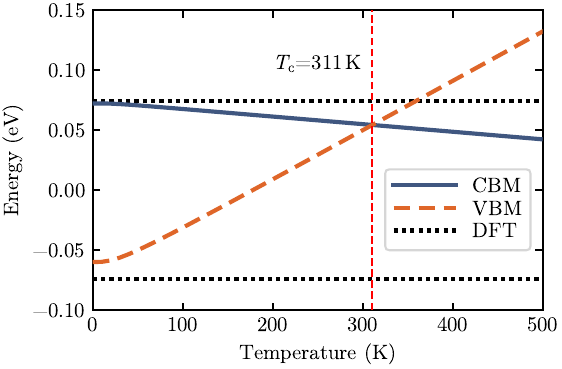}
\caption{
Temperature dependence of the VBM and CBM energies.
The VBM and CBM energies computed without EPC are indicated with the black horizontal dotted lines.
The critical temperature at which the band gap between the VBM and CBM vanishes is indicated with the red vertical dashed line.
}
\label{fig:S_t_gap}
\end{figure}

Figure~\ref{fig:S_t_gap} shows the temperature dependence of the valence band maximum (VBM) and the conduction band minimum (CBM) energies.
\bts{} is a direct-gap insulator with both the VBM and the CBM at the $\Gamma$ point, where the off-diagonal self-energy is zero due to the inversion symmetry of the crystal structure.
Therefore, the temperature dependence of the VBM and CBM states and the temperature dependence of the band gap can be computed without considering the off-diagonal self-energy as in Ref.~\cite{2016Antonius}.

The transition temperature we obtained, $T_{\rm c} = 311\,{\rm K}$, is a little higher than the transition temperature obtained in Ref.~\cite{2016Antonius}, which is around 240~K.
We believe that this discrepancy mainly originates from the fact that the band gap of our DFT calculation is a little larger than that of Ref.~\cite{2016Antonius}.
The difference in the band gap might be attributed to the use of different exchange-correlation functionals: we used the PBE functional~\cite{1996PerdewPBE} while the authors of Ref.~\cite{2016Antonius} used the revised PBE~\cite{1998ZhangrevPBE} functional.

\makeatletter\@input{xx.tex}\makeatother

\bibliography{main}